%% file: main.tex
\documentclass[conference]{IEEEtran}
\IEEEoverridecommandlockouts
\usepackage{cite}
\usepackage{amsmath,amssymb,amsfonts}
\usepackage{graphicx}
\usepackage{textcomp}
\usepackage{xcolor}
\usepackage{hyperref}

\usepackage{amsmath,amsfonts,amsthm}
\usepackage{algorithm}
\usepackage[noend]{algpseudocode}
\algrenewcommand{\algorithmiccomment}[1]{\hfill {\small /* #1 */}}

\usepackage{graphicx}
\usepackage{textcomp}
\usepackage{colortbl}
\usepackage{enumitem}
\usepackage{adjustbox}
\usepackage{multirow}
\usepackage{threeparttable}
\usepackage{subfigure}
\usepackage{mathtools}
\usepackage{makecell}
\usepackage{bm}
\usepackage{pifont}
\usepackage[capitalize,noabbrev]{cleveref}
\usepackage{xurl}

\usepackage{tikz}
\newcommand*\circled[1]{\tikz[baseline=(char.base)]{
            \node[shape=circle,draw,inner sep=1pt] (char) {#1};}}

\DeclarePairedDelimiter\abs{\lvert}{\rvert}
\DeclarePairedDelimiter{\ceil}{\lceil}{\rceil}

\numberwithin{equation}{subsection}

\def\BibTeX{{\rm B\kern-.05em{\sc i\kern-.025em b}\kern-.08em
    T\kern-.1667em\lower.7ex\hbox{E}\kern-.125emX}}
\begin{document}

\title{\huge{Enabling Homomorphic Analytical Operations on Compressed Scientific Data with Multi-stage Decompression}
}


\author{
\centering
\begin{tabular}{ccc}
Xuan Wu & Sheng Di & Tripti Agarwal \\
\textit{Oregon State University} & \textit{Argonne National Laboratory} & \textit{University of Utah} \\
Corvallis, USA & Lemont, USA & Salt Lake City, USA \\
wuxuan@oregonstate.edu & sdi1@anl.gov & tripti.agarwal@utah.edu \\
\\[-3pt]
Kai Zhao & Xin Liang\IEEEauthorrefmark{1}\thanks{*Corresponding author: Xin Liang, School of Electrical Engineering and Computer Science, Oregon State University, Corvallis, OR 97331.  The source code is available at \url{https://github.com/xuanwu02/homoApplication}.} & Franck Cappello \\
\textit{Florida State University} & \textit{Oregon State University} & \textit{Argonne National Laboratory} \\
Tallahassee, USA & Corvallis, USA & Lemont, USA \\
kai.zhao@fsu.edu & lianxin@oregonstate.edu & cappello@mcs.anl.gov \\
\end{tabular}
\vspace{-1em}
}

\maketitle

\begin{abstract}
Error-controlled lossy compressors have been widely used in scientific applications to reduce the unprecedented size of scientific data while keeping data distortion within a user-specified threshold.
While they significantly mitigate the pressure for data storage and transmission, they prolong the time to access the data because decompression is required to transform the binary compressed data into meaningful floating-point numbers.  
This incurs noticeable overhead for common analytical operations on scientific data that extract or derive useful information, because the time cost of the operations could be much lower than that of decompression. 
In this work, we design an error-controlled lossy compression and analytical framework that features multi-stage decompression and homomorphic analytical operation algorithms on intermediate decompressed data for reduced data access latency. Our contributions are threefold. (1) We abstract a generic compression pipeline with partial decompression to multiple intermediate data representations and implement four instances based on state-of-the-art high-throughput scientific data compressors. (2) We carefully design homomorphic algorithms to enable direct operations on intermediate decompressed data for three types of analytical operations on scientific data. (3) We evaluate our approach using five real-world scientific datasets. Experimental evaluations demonstrate that our method achieves significant speedups when performing analytical operations on compressed scientific data across all three targeted analytical operation types.
\end{abstract}



\input{tex/introduction}
\input{tex/related}

\input{tex/formulation}
\input{tex/method}
\input{tex/evaluation}
\input{tex/conclusion}

\section*{Acknowledgment}
The material was supported by the U.S. Department of Energy, Office of Science, and by DOE’s Advanced Scientific Research Computing Office (ASCR) under contract DE-AC02-06CH11357, and supported by the National Science Foundation under Grant No. OAC-2311875, OAC-2344717, OAC-2514036, OAC-2513768, OAC-2442627, and OIA-2327266.

\section*{AI-Generated Content Acknowledgement}
Portions of the text in this manuscript were refined using language-assistance tools such as ChatGPT (GPT-5). These tools were used solely to improve grammar, clarity, and readability. All conceptualization, data analysis, and scientific conclusions are entirely the work of the authors.

\bibliographystyle{IEEEtran}
\bibliography{sample}

\end{document}

%% file: tex/introduction.tex
\section{introduction}

Scientific applications running on cutting-edge high-performance computing systems generate vast amounts of data. 
Such scientific data consists of multiple high-dimensional floating-point arrays across various physical fields, overwhelming the underlying data management systems. 
For example, modern climate simulations generate 16 TB of data per second~\cite{foster2017computing}, while the typical I/O rate is 1 TB per second. 
This poses significant challenges for data storage and transmission, which become major bottlenecks in scientific computing.

Compression is regarded as one of the most promising ways to address data challenges, and it has been extensively used in database design~\cite{cockshott1998high, chang2004retrieving, taskiran2004vibe, deri2012tsdb} and analytical operation acceleration~\cite{chen2001query, arion2007xquec, zhang2018efficient}.
For scientific data management, error-controlled lossy compression reduces data size while enforcing user-specified error tolerance~\cite{cappello2019use}. 
It has significantly reduced data movement cost and has been widely adopted in climatology~\cite{baker2017toward}, cosmology~\cite{jin2021adaptive}, molecular dynamics~\cite{zhao2022mdz}, and quantum chemistry~\cite{gok2018pastri}. 

Although compression substantially reduces scientific data size, it increases data-access latency because decompression is required to recover values. 
This overhead can dominate many analytical operations that are computationally inexpensive. For example, regional minima and maxima for isosurface detection~\cite{lorensen1987marching}, as well as numerical differentiation for climate analysis~\cite{hieronymus2022algorithmic}, can be much cheaper than full decompression.

Recently, homomorphic compression, which allows for operations on compressed data without decompression, has been proposed as a novel concept in the data management community. 
It has been broadly applied in analyzing and processing text data~\cite{agarwal2015succinct, pan2021exploring, zhang2021g, guan2023homomorphic, zhang2022compressdb} in database management systems. 
Despite its success, limited research has focused on homomorphic compression for floating-point scientific data, which prolongs the time needed to obtain analytical results.

Designing effective homomorphic compression methods for scientific data is non-trivial. 
Scientific compressors typically apply multiple complex transforms, making end-to-end homomorphism difficult. In addition, scientific analytics rely on diverse arithmetic operations that differ substantially from those in text processing. Homomorphic methods must also match the compression efficiency and performance of existing compressors to keep pace with growing data volumes, while providing strict error control to preserve scientific fidelity.

In this paper, we present HSZ, a novel compression framework that accelerates analytical operations on scientific data without compromising the compression ratios and error controls. 
While following the same compression pipeline as state-of-the-art scientific data compressors, HSZ features four progressive decompression stages, with homomorphic algorithms to facilitate plausible analytical operations in each stage. 
Such design not only saves decompression time, but may also reduce operation time due to simplified data types and structures. 
We implement four new compressors based on two state-of-the-art high-performance error-controlled scientific data compressors, namely SZp~\cite{huang2024hzccl} and SZx~\cite{yu2022szx}, and enable fast execution of six analytical operations on scientific data from three categories. 
In summary, our contributions are as follows. 

\begin{itemize}
    \item We design a generic compression pipeline with partial decompression in multiple stages, allowing compressed data to be decompressed into four progressive stages: floating-point data, direct quantization integers, predicted quantization integers, and metadata, with increasing throughput. 
    We further adapt SZp and SZx, two error-controlled lossy compressors with leading compression and decompression throughput, to our framework and investigate their multidimensional variations to achieve higher compression ratios.
    \item We develop homomorphic algorithms to support a variety of analytical operations on the four decompression stages provided in our framework. 
    In particular, we develop tailored methods to perform the operations using intermediate data generated by partial decompression, taking advantage of their high decompression throughput and simple-to-operate data types and structures. 
    \item We evaluate our methods and compare them with state-of-the-art baselines using four real-world datasets. Experimental results demonstrate that our framework leads to up to 7315$\times$, 1.89$\times$, and 2.68$\times$ performance gains in performing analytical operations on scientific data related to statistical computation, numerical differentiation, and multivariate derivation, respectively. 
\end{itemize}

The rest of the paper is organized as follows. \cref{sec:related} reviews related work. \cref{sec:problem} formulates the problem and outlines the proposed framework. Sections~\ref{sec:compression} and~\ref{sec:query} describe the compression design and homomorphic analytical operations, respectively. \cref{sec:evaluation} presents evaluation results on real-world datasets, and \cref{sec:conclusion} concludes with a discussion of future work.

%% file: tex/related.tex
\section{related work}\label{sec:related}
In this section, we review state-of-the-art error-controlled lossy compressors for scientific data and existing homomorphic compression techniques that support operations on compressed data.

\subsection{Error-controlled Lossy Compressors}
The ever-increasing amount of data produced by scientific applications and instruments has made compression essential for efficient storage, management, and analysis.
Generic lossless compressors such as GZIP~\cite{gzip} and ZSTD~\cite{zstd}, achieve limited compression on floating-point data due to random mantissas, while traditional lossy methods (e.g., JPEG and JPEG2000~\cite{wallace1992jpeg,rabbani2002jpeg2000}) lack strict error guarantees and can compromise data fidelity. Error-controlled lossy compression addresses these limitations by providing both effective compression ratios and guaranteed error bounds.

Error-controlled lossy compressors are generally categorized as prediction-based~\cite{lindstrom2006fast, lakshminarasimhan2013isabela, sz17, sz18, zhao2021optimizing, liu2021exploring, liu2024high, wu2025enabling} and transform-based~\cite{lindstrom2014fixed, ballester2019tthresh, li2023sperr}, depending on how the original data is decorrelated to more compressible representations. 
FPZIP~\cite{lindstrom2006fast} is a typical prediction-based compressor that relies on a multidimensional Lorenzo predictor~\cite{ibarria2003out} for data decorrelation and arithmetic encoders~\cite{witten1987arithmetic} for residual encoding, and it provides guaranteed control on point-wise relative errors. 
Later, SZ~\cite{liang2022sz3} extends the design by adopting linear-scaling quantization~\cite{sz17} to enable absolute error control, and then explores more advanced prediction methods such as regression~\cite{sz18} and splines~\cite{zhao2021optimizing} to achieve better decorrelation efficiency and thus higher compression ratios. 
It also leverages Huffman encoding~\cite{huffman1952method} and existing lossless compressors to reduce the compressed data size further. 
ZFP~\cite{lindstrom2014fixed} is a popular transform-based compressor that leverages domain transforms for decorrelation. 
In particular, it divides the original data into dependent blocks, and utilizes a near-orthogonal transform to convert block data into coefficients, which are then encoded by an embedded encoding algorithm. 
After that, TTHRESH~\cite{ballester2019tthresh} investigates the use of singular value decomposition to achieve better compression ratios at the cost of slower compression and decompression throughput. 
Meanwhile, many other methods for error-controlled lossy compression have also emerged. 
For instance, MGARD~\cite{ainsworth2018multilevel, ainsworth2019multilevel} combines finite element analysis and wavelet theories to offer guaranteed error control for derived quantities, and several variations~\cite{liang2020toward, jiao2022toward, yan2023toposz, xia2024preserving, xia2025tspsz} of the aforementioned compressors have enabled error control on topological features. 

Although error-controlled lossy compressors have demonstrated high efficiency in reducing the size of scientific data and have been widely used for data archival, they unavoidably introduce extra costs for accessing the data. 
Because most of the compressed data is in binary format, decompression is required to transform it into meaningful floating-point numbers, resulting in noticeable overhead for post hoc data analytics. 
Furthermore, such overhead becomes significant when the analytical operations are cheaper than decompression, which is often the case for statistical computation and quantity derivation widely used in multiple scientific discipines. 

Recently, two lightweight error-controlled lossy compressors, SZx~\cite{yu2022szx} and SZp~\cite{huang2024hzccl}, have been proposed to provide fast data compression and decompression at the cost of reduced compression ratios. 
They feature simplified data decorrelation and encoding procedures, delivering very high throughput on a wide range of scientific datasets. 
While they can be used to reduce decompression overhead, they still suffer from the cost of full decompression, leading to suboptimal performance for analytical operations on scientific data.

\subsection{Homomorphic Compression}
The idea of homomorphic compression originates from homomorphic encryption~\cite{yi2014homomorphic, acar2018survey}, which represents an encryption scheme that allows for performing computable functions on encrypted data without decryption. 
Similarly, homomorphic compression is used to refer to specific compression techniques that permit computable operations on the compressed data without decompression. 

Homomorphic compression techniques~\cite{agarwal2015succinct, rodrigues2021clp, zhang2021tadoc, zhang2021g, pan2021exploring, zhang2022compressdb, guan2023homomorphic} have already been used in the data management community to reduce the latency of analytical operations to compressed data formats. 
In~\cite{agarwal2015succinct}, Succinct is proposed to support count and search of arbitrary strings as well as range and wildcard operations without storing indexes. 
In~\cite{zhang2021tadoc}, TADOC is developed to compress text data and perform text analytics, including word count and inverted index, without decompression, and it has been extended to GPU to achieve high throughput~\cite{zhang2021g}. 
In~\cite{zhang2022compressdb}, a new storage engine is designed to support data processing for databases without decompression. 
Later, an efficient text data management engine~\cite{guan2023homomorphic} is proposed to generalize homomorphic text data processing with more compression schemes and text analytics. 
Despite the success of these methods in managing text data, it is infeasible to apply them to scientific data, where floating-point data representations and the related operations require new compression mechanisms and homomorphic algorithms. 

Recently, several attempts have been made to extend homomorphic compression to floating-point data. 
Stochastic quantization has been leveraged to enable homomorphic aggregation of weights for neural networks during distributed training~\cite{li2024thc}. However, it suffers from limited compression ratios and unbounded errors, and can only be used for aggregation operations. 
SZOps~\cite{szops2024}, a variation of SZp~\cite{huang2024hzccl}, has been proposed to support scalar operations and statistical analytics, such as mean and variance.
However, it only considers one decompression level of one specific compressor, leading to suboptimal efficiency. 
Furthermore, it only targets simple operations, such as scalar addition and multiplication, that apply uniformly to the data, and its current design cannot handle complex analytical operations that derive new quantities from the original data. 

In this work, we propose HSZ, a generic compression and analytical framework that supports partial decompression and homomorphic analytical operations at multiple decompression stages. 
In particular, we design four intermediate decompression stages and develop specific homomorphic operations for three types of scientific analytics. 
We also implement four new compressors based on two leading high-throughput error-controlled lossy compressors, to accommodate use cases with diverse requirements for desired compression ratios and operation throughput.

%% file: tex/formulation.tex
\section{Problem Formulation and Overview}\label{sec:problem}
We formulate our research problem in this section, and provide an overview of the proposed compression framework. 

\subsection{Objective}
Generic error-controlled lossy compressors compress floating-point scientific data $\mathbf{d}=\{d_1, \dots, d_n\}$ into compressed format $\mathbf{c} = \mathcal{C}(\mathbf{d})$ with a user-specified error bound $\epsilon$ and store $\mathbf{c}$ for data archival or transfer. 
Decompression reconstructs $\mathbf{d'}=\mathcal{D}(\mathbf{c})$ such that $\abs{d'_i - d_i} \leq \epsilon$ for all $i$.
Let $\texttt{size}(\cdot)$ denote data size. The compression ratio is defined as $\texttt{size}(\mathbf{d})/\texttt{size}(\mathbf{c})$, which indicates the rate of size reduction.

Since compressed data $\mathbf{c}$ is stored in plain binary format, analytical operations require decompression to obtain the floating point representation $\mathbf{d'}=\mathcal{D}(\mathbf{c})$. Let $t_{\texttt{op}}(\cdot)$ denote the time to perform an operation \texttt{op}, and let $\mathcal{Q}$ be the target analytical operation. The total time to operate on compressed scientific data is therefore:
\vspace{-.5em}
\begin{equation*}
    t_{total} = t_{\mathcal{D}}(\mathbf{c}) + t_{\mathcal{Q}}(\mathbf{d'}).
\vspace{-.5em}
\end{equation*}

\emph{The goal of this work is to investigate the use of partial decompression and homomorphic algorithms to reduce the latency $t_{total}$ for performing analytical operations on compressed scientific data.}
We address this by enabling partial decompression for error-controlled data compressors in multiple stages and designing homomorphic algorithms to perform analytical operations using the intermediate data generated by that. 
Formally, our homomorphic compressor compresses original data $\mathbf{d}$ into compressed format $\mathbf{c} = \mathcal{C}(\mathbf{d})$ with user-specified error bound $\epsilon$ similar to existing scientific data compressors, but our decompression procedure is decomposed to $m$ progressive stages $\{\mathcal{D}_1, \dots \mathcal{D}_m\}$ such that $\mathbf{d'} = \mathcal{D}(\mathbf{c}) = \mathcal{D}_m(\dots \mathcal{D}_1(\mathbf{c}))$.
Let $\mathbf{d}^{(i)} = \mathcal{D}_i(\dots \mathcal{D}_1(\mathbf{c}))$ denote the intermediate data representation after applying $i$ stages. 
Let $\mathcal{Q}_i$ represent the homomorphic algorithm that executes operation $\mathcal{Q}$ on $\mathbf{d}^{(i)}$.
The minimal operation execution time now becomes: 
\vspace{-.5em}
\begin{equation*}
    t_{total} = \min_i t_{\mathcal{D}_i\dots \mathcal{D}_1}(\mathbf{c}) + t_{\mathcal{Q}_i}(\mathbf{d}^{(i)}).
\vspace{-.5em}
\end{equation*}
This reduces to generic error-controlled scientific data compression when $m=1$, but challenges arise when $m>1$. 
In particular, we need to complete the following three tasks to solve the optimization problem formulated above: (1) enable multi-stage decompression in error-controlled scientific data compressors; (2) develop homomorphic methods $\mathcal{Q}_i$ to perform analytical operations on intermediate data representations; and (3) identify the proper stage to perform decompression and the operations. 
We address the first two tasks in Section~\ref{sec:compression} and Section~\ref{sec:query}, respectively, and provide evaluation results for the last task in Section~\ref{sec:evaluation}.

\subsection{Target Operations on Scientific Data}
In this paragraph, we introduce the targeting analytical operations on scientific data in this paper, which fall into three categories as outlined below. For each category, we present two representative operations as a proof of concept, and the design of our pipeline is inherently extensible to a broader class of operations. Specifically, operations that can be expressed through local stencils, linear combinations, or blockwise aggregations can be similarly supported by adopting the same underlying principles.

\paragraph{\textbf{Statistical computation}} Obtaining statistics such as mean, standard deviation, minimum/maximum from scientific data is essential for understanding its tendency and dispersion. 
While directly storing the statistics is feasible during compression, the overhead becomes prohibitive when regional statistics are needed, which is often the case in scientific data analytics. 

Regarding statistical computation, we primarily investigate analytical operations for the mean and standard deviation, which are the most widely used statistics in scientific data analytics. 
For instance, the mean of the data field ``dark\_matter\_density'' in cosmological simulation is used as an essential indicator for identifying halos~\cite{almgren2013nyx}. 
Given a scalar data field $\mathbf{d}$, its mean $\mu$ and standard deviation $\sigma$ are defined as follows: 
\begin{equation}
    \mu = \frac{\sum_{i=1}^{n}d_i}{N},\; \sigma = \sqrt\frac{\sum_{i=1}^{n}(d_i-\mu)^2}{N-1}.
    \label{eq:var}
\end{equation}

\paragraph{\textbf{Numerical differentiation}} Scientific applications usually apply numerical differentiation to data to obtain additional information for better understanding and visualization. 
Compressing and storing this additional information separately often results in unnecessary storage overhead, as its size can be the same as (or even larger than) that of the original data. 

As for numerical differentiation, we mainly support analytical operations for derivatives and the Laplacian, which are essential derived quantities in a broad range of scientific applications. 
For example, derivatives are commonly used in climate analysis to detect the sensitivity of state variables~\cite{hieronymus2022algorithmic}, 
and discrete Laplacian stencils can serve as implicit spatial filters on climate-model ocean velocity and temperature fields~\cite{danilov2024extracting}, enabling the isolation of mesoscale and basin-scale structures without relying on global transforms.
Given a 2D scalar data field $\mathbf{d}$, its derivative vector $\nabla d_{i,j}$ and Laplacian $\Delta d_{i,j}$ at coordinate $(i, j)$ can be approximated by finite element methods as: 
\begin{align}
    \nabla d_{i,j} & = (\delta_x d_{i,j}, \delta_y d_{i,j}) = (\frac{d_{i+1,j} - d_{i-1,j}}{2}, \frac{d_{i,j+1} - d_{i,j-1}}{2}) ,
    \label{eq:derivatives}\\
    \Delta d_{i,j} &= d_{i-1,j} +  d_{i+1,j} +  d_{i,j-1} +  d_{i,j+1} - 4d_{i,j}. 
    \label{eq:laplacian}
\end{align}

\paragraph{\textbf{Multivariate derivation}} In addition to derived information from a single variable, scientific data analytics may require more complex derived information from multiple variables. 
Again, it is inefficient to compress and store this information separately, due to prohibitive storage costs. 

As for the multivariate derivation, we mainly focus on the divergence and curl, two multivariate derived quantities with multiple uses in vector fields. 
For instance, the magnetic field in fusion simulation is required to be divergence-free to avoid the presence of magnetic monopoles~\cite{chang2017fusion}, so verifying its divergence is a necessary step.
Meanwhile, the wind stress curl plays a critical role in driving and modulating oceanic vorticity and mesoscale circulation~\cite{rai2025wind}, therefore computing its curl is essential for diagnosing energy transfer and dynamic balance in coupled ocean–atmosphere systems.

Formally, given a 2D vector field $\mathbf{F} = (\mathbf{u}, \mathbf{v})$, its divergence $\nabla \cdot \mathbf{F}$ and curl $\nabla \times \mathbf{F}$ are defined as:
\begin{align}
    (\nabla\!\cdot\!\mathbf{F})_{i,j}
    &= \delta_x u_{i,j} + \delta_y v_{i,j},
    \label{eq:divergence}\\[4pt]
    (\nabla\!\times\!\mathbf{F})_{i,j} &= \delta_y u_{i,j} - \delta_x v_{i,j}
    \label{eq:curl}
\end{align}
where the $\delta_x$ and $\delta_y$ are numerical differentiation operators similar to those defined in Eq.~\ref{eq:derivatives}.

\subsection{Design Overview}\label{sec:overview}
\begin{figure}[ht]
	\includegraphics[width=0.95\columnwidth]{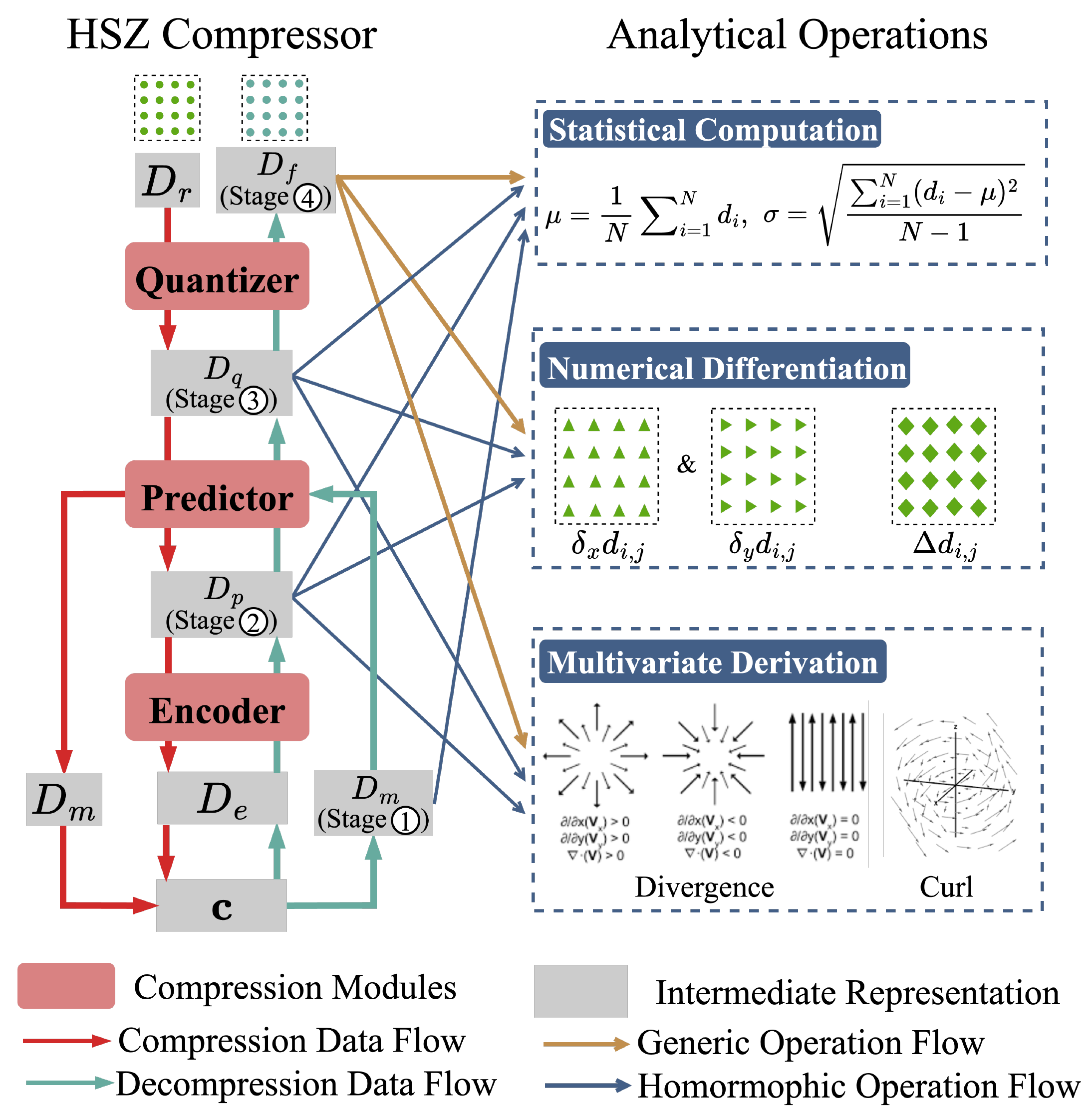}
	\centering
	\vspace{-1em}
	\caption{Overview of HSZ compression and homomorphic analytical operation pipelines. $D_m$ (metadata), $D_p$ (decorrelated data), $D_q$ (quantized data), $D_f$ (fully decompressed data) represent intermediate data representations in different decompression stages. }\label{fig:overview}
	\vspace{-1em}
\end{figure}

\vspace{-.3em}
We present an overview of the proposed HSZ framework in Figure~\ref{fig:overview}, which aims to accelerate analytical operation on scientifc data with partial decompression and homomorphic algorithms. The left side depicts the compression pipeline, and the right side illustrates the operation flow. 
Red boxes denote key modules for compression, and gray boxes represent intermediate data in different compression and decompression stages, with the detailed descriptions in Table~\ref{tab:notations}. 
Traditional analytical operation pipelines on compressed data require full decompression, resulting in substantial time overhead. In contrast, HSZ enables homomorphic operation execution by operating directly on intermediate data at specific stages of decompression, eliminating the need for full decompression. We further investigate the most suitable decompression stages for various operations through rigorous analysis and thorough evaluation.
In the following sections, we introduce the compression pipeline and multi-stage decompression in HSZ, followed by the homomorphic analytical operations. 

\vspace{-.2em}

\vspace{-1.0em}
\begin{table}[ht]
    \centering
    \footnotesize
    \setlength{\tabcolsep}{4pt}
    \caption{Notations for intermediate data representation in different decompression stages.}
    \vspace{-1em}
    \begin{tabular}{|c|l|l|l|}
        \hline
        \textbf{Notation} & \textbf{Stage} & \textbf{Type} & \textbf{Description}  \\
        \hline
        $D_r$ & - & Floaing-point & Original data. \\
        \hline
        $\mathbf{c}$ & - & Binary & Compressed data. \\
        \hline
        $D_f = \{d'_i\}$ & \circled{4} &  Floaing-point & Fully decompressed data.\\
        \hline
        $D_q = \{q_i\}$ & \circled{3} & Integer & Quantized data.\\
        \hline
        $D_p = \{p_i\}$ & \circled{2} & Integer & Decorrelated data. \\
        \hline
        $D_m = \{M_i\}$ & \circled{1} & Integer/Floating-point & Metadata. \\
        \hline
    \end{tabular}
    \label{tab:notations}
    \vspace{-1em}
\end{table}

%% file: tex/method.tex
\section{Error-controlled Compression with Partial Decompression}\label{sec:compression}

We follow a generic compression pipeline~\cite{liang2022sz3} to enable partial decompression in HSZ and demonstrate how to implement it using two high-throughput scientific data compressors, SZp~\cite{huang2024hzccl} and SZx~\cite{yu2022szx}. 
We use SZp and SZx due to their high performance, and the methodology directly generalizes to other compressors with the same decorrelation paradigm (e.g., SZ-1.4~\cite{sz17} and cuSZ~\cite{tian2020cusz}). 

Since both SZp and SZx treat the original data as a 1D array regardless of their original dimensions, we describe the procedure using 1D data with $n$ points as an example. 
The high-level compression algorithm is presented in Algorithm~\ref{alg:compression}, following the prediction-based compression pipeline in~\cite{liang2022sz3}. 
The key steps are detailed below. 

\textit{Quantization (lines 1-3):}
Let $d_{i}$ be the $i$-th original data point. A linear-scaling quantizer maps the floating-point value to an integer $q_{i}$ with a user-specified error bound $\epsilon$ using the formula $q_{i} = \texttt{round}(d_{i}/(2\epsilon))$, where $\texttt{round}(\cdot)$ is the rounding operator.
During decompression, the decompressed data is recovered by $d'_{i} = 2q_{i}\epsilon$, which ensures $\abs{d_{i} - d'_{i}} \leq \epsilon$.
This step is applied to all the data points, which maps the original floating-point data array $D_r$ to a quantization integer array $D_q$. 

\textit{Data Partition (line 4):} 
Denoting the block size $S$ as a tunable parameter, the original data is uniformly divided into $\ceil{\frac{n}{S}}$ blocks. 

\textit{Metadata Collection (line 6):} 
This step collects metadata $D_m$ that will be used for decorrelation. 
In SZp, the first quantized data in a block is stored as the metadata; in SZx, the blockwise means is stored to enable mean-based prediction. 

\textit{Decorrelation (line 9):} 
Decorrelation reduces the entropy of the quantized data, leading to higher compressibility and, consequently, higher compression ratios. 
In SZp, any quantized data $q_{i}$ (except the first one in the block which is stored as metadata) is predicted by $q_{i-1}$ to generate the decorrelated data $p_{i} = q_{i} - q_{i-1}$.
In SZx, this is done by using the mean of the maximum and minimum inside a block, i.e., $p_{i} = q_{i} - \texttt{mean}(\texttt{max}(B_{i_0}), \texttt{min}(B_{i_0}))$, where $B_{i_0}$ denotes the blocking containing the $i$-th data point.  
This stage transforms the quantization index array $D_q$ to a decorrelated data array $D_p$ with less entropy.

\textit{Encoding (line 11):}
Encoding is the stage for actual size reduction in the compression pipeline. 
While we adopt the blockwise fixed-rate encoding method used in SZp for high throughput, our framework can accommodate any other encoding methods, such as Huffman coding~\cite{huffman1952method} and run-length encoding~\cite{golomb1966run}.
In particular, this method first traverses decorrelated data $D_q$ in a block, to record the signs and determine an appropriate number of bits to encode the data based on the maximal absolute value.
After that, the sign bits are put together, and the values are recorded in order using their binary representations with the designated number of bits.

\setlength{\textfloatsep}{2mm}
\begin{algorithm}[t]
\caption{HSZ compression} \label{alg:compression} \footnotesize
\begin{flushleft}
\textbf{Input}: Input data $\mathbf{D}$, error bound $\epsilon$\\
\textbf{Output}: Compressed data $\mathbf{c}$
\end{flushleft}
\begin{algorithmic} [1]
\For{$i = 1 \to N$} \Comment{Perform quantization (obtain $D_q$ and $D_f$)}
    \State $\mathbf{D}_q[i] \gets \texttt{quantize}(\mathbf{D}[i])$
\EndFor
\State $\{B_1, \dots, B_{n_b}\}\gets$ $\texttt{partition}(\mathbf{D})$\Comment{partition data to blocks}
\For{$ = 1 \to n_b$} \Comment{Iterate each block}
    \State $M_i \gets \texttt{getMetadata}(\mathbf{D}, B_i)$ \Comment{Extract metadata (obtain $D_m$)}
    \For {$x \in B_i$}
        \State $i \gets \texttt{index}(x)$
        \State $D_p[i] \gets \texttt{decorrelate}(\mathbf{D_q}, M_i)$ \Comment{Perform decorrelation (obtain $D_p$)}
    \EndFor
    \State $c_i \gets \texttt{encode}(D_p[B_i])$ \Comment{Encode into binary format (obtain $D_e$)}
\EndFor
\State \textbf{return} $\{M_i\} \cup \{c_i\}$
\end{algorithmic}
\end{algorithm}

\textbf{Multi-stage Decompression}:
The above compression pipeline allows partial decompression for obtaining different intermediate representations on a block-wise basis, as detailed in Algorithm~\ref{alg:decompression}.
In particular, we can choose to decompress the data to four different stages: metadata $D_m$ (line 3), decorrelated data $D_p$ (lines 5-6), quantized data $D_q$ (lines 8-12), and decompressed data $D_f$ (lines 14-18). 
The $\texttt{decode}(\cdot)$ operation inverts encoding by separately decoding the sign and value arrays and combining them to form decorrelated data $D_p$. The $\texttt{recorrelate}(\cdot)$ operation reverses decorrelation, transforming $D_p$ into quantized data $D_q$ using the same formula. The $\texttt{dequantize}(\cdot)$ operation recovers floating point values from $D_q$ by multiplying by $2\epsilon$.
Decompression to lower level stages requires less computation and typically yields higher throughput, but it demands more sophisticated homomorphic algorithms on intermediate representations. Higher level representations incur greater decompression cost but allow simpler homomorphic operations. When fully decompressed data $D_f$ (stage \circled{4}) is used, the process reduces to standard analytical operations on floating point data.

\algnewcommand\algorithmicswitch{\textbf{switch}}
\algnewcommand\algorithmiccase{\textbf{case}}
\algnewcommand\algorithmicassert{\texttt{assert}}
\algnewcommand\Assert[1]{\State \algorithmicassert(#1)}%
\algdef{SE}[SWITCH]{Switch}{EndSwitch}[1]{\algorithmicswitch\ #1\ \algorithmicdo}{\algorithmicend\ \algorithmicswitch}%
\algdef{SE}[CASE]{Case}{EndCase}[1]{\algorithmiccase\ #1}{\algorithmicend\ \algorithmiccase}%
\algtext*{EndSwitch}%
\algtext*{EndCase}%
\begin{algorithm}[t]
\caption{HSZ partial decompression in block} \label{alg:decompression} \footnotesize
\renewcommand{\algorithmiccomment}[1]{ /*#1*/}
\begin{flushleft}
\textbf{Input}: Metadata $M_i$, encoded data $c_i$, block indices $B_i$, decompression level $l$\\
\textbf{Output}: Intermediate data representation
\end{flushleft}
\begin{algorithmic} [1]
\Switch{$l$}
\Case{$0:$}
    \State \textbf{return} $M_i$ \Comment{Directly return metadata}
\EndCase
\Case{$1:$}
    \State $D_p[B_i] \gets \texttt{decode}(\mathbf{c_i})$ \Comment{Perform decoding}
    \State \textbf{return} $D_p[B_i]$ \Comment{Return decorrelated data}
\EndCase
\Case{$2:$}
    \State $D_p[B_i] \gets \texttt{decode}(\mathbf{c_i})$
    \For {$x \in B_i$}
        \State $D_q[x] \gets \texttt{recorrelate}(\mathbf{D_p}[x], M_i)$ \Comment{Recover quantized data}
    \EndFor    
    \State \textbf{return} $D_q[B_i]$ \Comment{Return quantized data}
\EndCase
\Case{$3:$}
    \State $D_p[B_i] \gets \texttt{decode}(\mathbf{c_i})$
    \For {$x \in B_i$}
        \State $d_q \gets \texttt{recorrelate}(\mathbf{D_p}[x], M_i)$ 
        \State $D_f[x] \gets \texttt{dequantize}(d_q)$ \Comment{Restore decompressed data}
    \EndFor    
    \State \textbf{return} $D_f[B_i]$ \Comment{Return decompressed data}
\EndCase
\EndSwitch
\end{algorithmic}
\end{algorithm}

We then leverage this framework to implement HSZp and HSZx, the homomorphic versions of SZp and SZx that feature multi-stage partial decompression. 
We further extend HSZp and HSZx to their multi-dimensional variations, which yield higher compression ratios at the cost of lower throughput. 
Table~\ref{tab:compressors} summarizes the characteristics of the four compressors that we implement in HSZ, and we detail the implementations as follows. 

\begin{table}[ht]
\centering
\caption{Characteristics of the four compressors}\label{tab:compressors}
\resizebox{\columnwidth}{!}{%
\begin{tabular}{|l|c|c|c|c| }
\hline 
Compressor & \makecell{Compression\\Ratios} & \makecell{Decompression\\Speed} & \makecell{Decompression\\Stages} \\ \hline
HSZp & Medium &  High  & \circled{2}\circled{3}\circled{4} \\\hline 
HSZp-nd &  High &  Low & \circled{2}\circled{3}\circled{4} \\\hline
HSZx & Low & High & \circled{1}\circled{2}\circled{3}\circled{4} \\\hline
HSZx-nd & High & Medium & \circled{1}\circled{2}\circled{3}\circled{4} \\\hline  
\end{tabular}
}
\end{table}

\textbf{HSZp:} We adopt the same compression pipeline as that in SZp, but revise the decompression procedure to make it multi-stage in HSZp. 
To avoid complex boundary handling, we also modify the decorrelation method to allow inter block prediction.
In particular, HSZp performs the decorrelation $p_{i} = q_{i} - q_{i-1}$ for any $i \neq 0$, while SZp stores any starting data in a block as metadata for block-independent decorrelation. 

\textbf{HSZp-nd:} 
We extend HSZp to HSZp-nd by leveraging multidimensional data partition and decorrelation, 
and use multidimensional Lorenzo prediction detailed in~\cite{ibarria2003out}. 
In particular, the decorrelation computes $p_{i,j} = q_{i,j} - q_{i,j-1} - q_{i-1,j} + q_{i-1,j-1}$ for 2D data and $p_{i,j,k} = q_{i,j,k} - q_{i,j,k-1,} - q_{i,j-1,k} - q_{i-1,j,k} + q_{i,j-1,k-1} + q_{i-1,j,k-1} + q_{i-1,j-1,k} - q_{i-1,j-1,k-1}$ for 3D data.

\textbf{HSZx:} We implement HSZx by adding the quantization stage to SZx and slightly changing its decorrelation method.
Specifically, we use the mean of all data in the block (instead of the mean of minimum and maximum) to facilitate fast computation of mean-related operations. 
The decorrelation becomes $p_{i} = q_{i} - \texttt{mean}(B_{i_0})$ where $B_{i_0}$ is the block containing the $i$-th data point. 

\textbf{HSZx-nd:} Similar to HSZp-nd, we extend HSZx to HSZx-nd with multidimensional data partition and decorrelation. 
Data partition is the same as that in HSZp-nd, and the decorrelation is performed using the mean of the multidimensional data block, i.e., $p_{i,j} = q_{i,j} - \texttt{mean}(B_{i_0j_0})$ for 2D data where $B_{i_0j_0}$ is the multidimensional data block containing the current data point. 
We illustrate the HSZx-nd approach using a small $2\times4$ example $Y = \begin{bmatrix}\begin{smallmatrix} 1.2 & 1.5 & -2.3 & -2.5 \\ 2.5 & -1.0 & 2.0 & 1.7\end{smallmatrix}\end{bmatrix}$. With an absolute error bound of 0.1, the quantized data is $Y_q = \begin{bmatrix}\begin{smallmatrix} 6 & 8 & -11 & -12\\ 13 & -5 & 10 & 9 \end{smallmatrix}\end{bmatrix}$. Using a $2\times2$ block size, $Y_q$ is partitioned into two blocks with integer means $M_1=5$ and $M_2=-1$. After block-wise decorrelation, we obtain $Y_p = \begin{bmatrix}\begin{smallmatrix} 1 & 3 & -10 & -11\\ 8 & -10 & 11 & 10 \end{smallmatrix}\end{bmatrix}$.

\section{Homomorphic analytical operation}\label{sec:query}
With multi stage decompression, we present the analytical operation pipeline in \cref{alg:query}.
For simplicity, we assume that each block operation depends only on its preceding and succeeding blocks, which generalizes to all target operations. To support large scientific datasets, we enable out of core execution using a sliding window.
Specifically, we first perform partial decompression to the designated stage using the first two data blocks (lines 3-4), and perform the operation on the first data block (line 5).
We then iterate over the remaining blocks (except the last one) to decompress the next data block and perform the operation (lines 6-9). 
Finally, we process the last data block (line 10).
This approach incurs minimal memory overhead since at most three data blocks are processed simultaneously, effectively handling cases where the entire dataset is too large to fit into memory. 

\begin{algorithm}[t]
\caption{Homomorphic analytical operation pipeline} \label{alg:query} \footnotesize
\renewcommand{\algorithmiccomment}[1]{ /*#1*/}
\begin{flushleft}
\textbf{Input}: Compressed data $\mathbf{c}$, decompression stage $l$, homomorphic operation algorithm\\
\textbf{Output}: Analytical operation result $\mathcal{R}$ 
\end{flushleft}
\begin{algorithmic} [1]
\State $\{M_i\}, \{c_i\} \gets \mathbf{c}$
\State $\{B_1, \dots, B_{n_b}\}\gets$ $\texttt{partition}(\mathbf{D})$\Comment{get data partitions}
\State $D_1 \gets \texttt{partialDecompress}(M_1, c_1, B_1, l)$
\State $D_2 \gets \texttt{partialDecompress}(M_2, c_2, B_2, l)$
\State $\mathcal{R}_{2} \gets \texttt{homomorphicAnaOp}(D_1, D_2)$
\For{$i = 2 \to n_b - 1$} \Comment{iterate each block}
    \State $D_{i+1} \gets \texttt{partialDecompress}(M_{i+1}, c_{i+1}, B_{i+1}, l)$
    \State $\mathcal{R}_i \gets \texttt{homomorphicAnaOp}(D_{i-1}, D_{i}, D_{i+1})$
\EndFor
\State $\mathcal{R}_{n_b} \gets \texttt{homomorphicAnaOp}(D_{n_b-1}, D_{n_b})$
\State return $\{\mathcal{R}_i\}$
\end{algorithmic}
\end{algorithm}

A critical problem in the analytical operation pipeline is enabling homomorphic operations on intermediate data representations. We develop homomorphic algorithms for the target operations using three representations, namely metadata $D_m$ (stage \circled{1}), decorrelated data $D_p$ (stage \circled{2}) and quantized data $D_q$ (stage \circled{3}) under the proposed compressors.
We omit the analysis for decompressed data $D_f$ in stage \circled{4} where the operations directly apply. 
For HSZp-nd and HSZx-nd compressors, we present 2D derivations and the approach generalizes naturally to higher dimensions.

\vspace{-0.2em}
\subsection{Statistical Computation}
\subsubsection{Mean} 
The homomorphic mean value computation can be completed in stages \circled{2}\circled{3} for SZp-based compressors and stages \circled{1}\circled{2}\circled{3} for SZx-based compressors, as detailed below. 

\circled{1}\textit{ Operation with $D_m$}:
The metadata that HSZx and HSZx-nd store provides a ultra-fast approach for computing the mean. With the easily accessible mean quantization index of each block, we can obtain the overall mean value by
\vspace{-.5em}
\begin{equation}
    \mu = \Big (\frac{1}{N}\sum_{b=1}^{n_b} M_b\cdot S_b\Big )
    \cdot 2\epsilon,
    \label{eq:mean_m}
\vspace{-.5em}
\end{equation}
where $M_b$ and $S_b$ are the quantized value of the mean in the block and the size of the block, respectively.

\circled{2}\textit{ Operation with $D_p$}:
For analytical operations that require pointwise information, the objective of operating on $D_p$ is to obtain this information without explicitly reconstructing $D_q$, which is often computationally expensive.

In HSZx and HSZx-nd, within block $b$, $D_p$ is formed by subtracting $M_b$ from the quantization index of each data point, leading to the following formulation for mean computation: 
$
\mu = \Big [ \frac{1}{N}\sum_{b=1}^{n_b} \Big ( \sum_{p\in D_p[B_b]} p + M_b\cdot S_b \Big ) \Big ]\cdot 2\epsilon.
$

In HSZp, within block $b$, the relation between $D_q$ and $D_p$ follows $q_i = p_i + q_{i-1}$, which implies $q_i = \sum_{t=1}^i p_t$. Consequently, we have $\sum_{i=1}^{S_b}q_i = \sum_{i=1}^{S_b}\sum_{t=1}^i p_t = \sum_{i=1}^{S_b} (S_b-i+1)\cdot p_i$. The mean is therefore computed as
$
\mu = \Big [ \frac{1}{N}\sum_{b=1}^{n_b} \Big ( \sum_{i=1}^{S_b} (S_b-i+1)\cdot p_i \Big ) \Big ] \cdot 2\epsilon.
$

In HSZp-2d, the decorrelated data is computed as $p_{i,j} = q_{i,j}-q_{i,j-1} - q_{i-1,j} + q_{i-1,j-1}$, giving the formula for computing $q_{i,j}$ using $D_p$: $q_{i,j} = \sum_{s=1}^i\sum_{t=1}^j p_{s,t}$. With this, we can immediately derive the mean calculation formula
$
\mu = \Big[\frac{1}{N}\sum_{s=1}^{n_1}\sum_{t=1}^{n_2} (n_1-s+1)(n_2-t+1) p_{s,t}\Big]\cdot2\epsilon.
$

\circled{3}\textit{ Operation with $D_q$}:
Because $D_q$ differs from $D_f$ by $2\epsilon$ for each data point, computing the mean using $D_p$ is straightforward:
$
\mu = \frac{1}{N}\Big(\sum_{q\in D_q} q\Big) \cdot 2\epsilon.
$

\subsubsection{Standard Deviation}
The homomorphic standard deviation computation can be completed in stages \circled{2}\circled{3} for all the compressors, as described in the following text.

\circled{2}\textit{ Operation with $D_p$}:
In HSZx and HSZx-nd, we have the integer mean $\tilde\mu = \texttt{round}(\frac{1}{N}\sum_{b=1}^{n_b} M_b\cdot S_b)$ according to Eq.~\ref{eq:mean_m}. 
Within the $b$-th block, the decorrelated data $D_p[B_b]$ stores the difference between each data point and the block mean, i.e., $p_i = q_i - M_b$. 
With this information, we can break down any single difference term of Eq.~\ref{eq:var} into $(q_i - M_b) + (M_b - \tilde\mu)$, and compute the standard deviation by:
$
\sigma
= \sqrt\frac{\sum_{b=1}^{n_b}\sum_{p\in D_p[B_b]} (p + M_b - \tilde\mu)^2}{N-1} \cdot 2\epsilon.
$
This computation is further simplified for constant blocks where $(p - M_b) = 0$ for all $p\in D_p[B_b]$.
Continuing the example of $Y$ in \cref{sec:compression}, we obtain $\tilde{\mu}=\texttt{round}(\frac{1}{8}(5\times4 + (-1)\times4)) = 2$ and $\sigma_{Y_p} = \sqrt{[(1+5-2)^2+\cdots+(10+(-1)-2)^2]/(8-1)}\cdot 0.2 = 2.000$.

In HSZp, Since it decorrelates data as $p_i = q_i - q_{i-1}$, each $q_i$ can be reconstructed as $q_i = \sum_{j=1}^i p_j$. During block processing, we maintain a scalar accumulator $\eta = \sum_{j=1}^i p_j$, updated for each element. Two additional scalar buffers, $s_1$ and $s_2$, accumulate $\sum q$ and $\sum q^2$ via $s_1\gets s_1 +\eta$ and $s_2\gets s_2 +\eta^2$. After processing all blocks, the standard deviation is computed as $\sigma = \sqrt{(s_2 - \frac{1}{N}s_1^2) / (N-1)} \cdot 2\epsilon$.

In HSZp-2d, \cref{alg:var_p_2d} converts the 2D Lorenzo reconstruction of $q_{i,j}$ into a simple 1D prefix-sum procedure, enabling direct accumulation of $\sum q$ and $\sum q^2$ without explicit data recovery. The 2D Lorenzo predictor satisfies 
$
q_{i,j} = q_{i-1,j} + \sum_{t\le j}p_{i,t},
$
revealing a row-wise dependency. We exploit this by maintaining a column buffer \texttt{colSum} of length $n_2$ to store the previously reconstructed row and a scalar \texttt{prefSum} to accumulate prefix sums of the current row. After processing row $i$, \texttt{colSum}$[j]$ equals $q_{i,j}$, and $q_{i,j}$ and $q_{i,j}^2$ have been added to $s_1$ and $s_2$ (line 8-9). For row $i+1$, \texttt{prefSum} accumulates $p_{i+1,j}$ (line 6), and is then added to \texttt{colSum}$[j]$ to reconstruct $q_{i+1,j}$, which is again accumulated into $s_1$ and $s_2$.

\begin{algorithm}[t]
\caption{HSZp-2d Standard Deviation (with $D_p$)} \label{alg:var_p_2d} \footnotesize
\renewcommand{\algorithmiccomment}[1]{ /*#1*/}
\begin{flushleft}
\textbf{Input}: Dimensions $n_1, n_2$, decorrelated data $D_p$, error bound $\epsilon$\\
\textbf{Output}: Standard Deviation $\sigma$
\end{flushleft}
\begin{algorithmic} [1]
\State allocMemory(\texttt{colSum}, $n_2$)
\State $s_1 \gets 0, s_2 \gets 0$\Comment{$s_1$ for $\sum q$, $s_2$ for $\sum q^2$}
\For{$i=1 \to n_1$}
    \State \texttt{prefSum}$\gets 0$
    \State $A_i \gets D_p[i]$\Comment{$A_i$ contains decorrelated data of row $i$}
    \For{$j=1 \to n_2$}
        \State \texttt{prefSum} += $A_i[j]$
        \State \texttt{colSum}$[j]$ += \texttt{prefSum}
        \State $s_1$ += \texttt{colSum}$[j]$
        \State $s_2$ += \texttt{colSum}$[j]$ $\cdot$ \texttt{colSum}$[j]$
    \EndFor
\EndFor
\State $\sigma \gets \sqrt{(s_2 - \frac{1}{N}s_1\cdot s_1) / (N-1)} \cdot 2\epsilon$
\State return $\sigma$
\end{algorithmic}
\end{algorithm}

\circled{3}\textit{ Operation with $D_q$}:
All of the four compressors use the same formula for the calculation of $\sigma$ as given below,
\vspace{-.5em}
\begin{equation}
    \sigma = \sqrt\frac{\sum_{q\in D_q} q^2 - \frac{1}{N}\left(\sum_{q\in D_q} q\right)^2}{N-1}\cdot2\epsilon. 
    \label{eq:var_q}
\end{equation}

\subsection{Numerical Differentiation}\label{sec:numerical}
HSZx and HSZp do not support homomorphic numerical differentiation at the decorrelated-data level (stage \circled{2}) because their partitioning does not preserve multidimensional layout.
We focus on homomorphic differentiation using $D_p$ only for HSZp-nd and HSZx-nd.

\subsubsection{Derivatives} As mentioned above, the homomorphic derivatives computation can be completed in stages \circled{2}\circled{3} for multidimensional compressors (HSZp-nd and HSZx-nd) and in stage \circled{3} for 1D compressors (HSZp and HSZx).

\circled{2}\textit{ Operation with $D_p$}:
In HSZx-2d, each block is decorrelated using a universal predictor (the block mean), so for any two points within the same block, $p_{i,j}-p_{i',j'} = q_{i,j}-q_{i',j'}$. For border points, differences in block means must be incorporated.
For example, when computing $\nabla d_{i,j}$ at the left boundary of block $b$, the left neighbor block $b-1$ is involved, yielding $\delta_y d_{i,j} = [p_{i,j+1} - p_{i,j-1} + (M_b - M_{b-1})]\cdot\epsilon$.
Similar adjustments apply to $\delta_x d_{i,j}$ at the top and bottom boundaries. Formally, for a block with integer mean $\mu_c$ and and neighboring block means $\mu_l, \mu_r, \mu_t, \mu_b$, we partition its indices into interior positions $\Omega_{in}$ and boundary sets $\Omega_l, \Omega_r, \Omega_t, \Omega_b$ corresponding to the left, right, top, and bottom borders, respectively.
For the \(x\)-direction,
$
\delta_x d_{i,j}
= \big(p_{i+1,j}-p_{i-1,j} + \Delta_x(i,j)\big)\cdot\epsilon,
$
where \(\Delta_x(i,j)=0\) for \((i,j)\in\Omega_{\mathrm{in}}\),
\(\Delta_x(i,j)=\mu_c-\mu_t\) for \((i,j)\in\Omega_t\),
and \(\Delta_x(i,j)=\mu_b-\mu_c\) for \((i,j)\in\Omega_b\).
For the \(y\)-direction,
$
\delta_y d_{i,j}
= \big(p_{i,j+1}-p_{i,j-1} + \Delta_y(i,j)\big)\cdot\epsilon,
$
where \(\Delta_y(i,j)=0\) for \((i,j)\in\Omega_{\mathrm{in}}\),
\(\Delta_y(i,j)=\mu_c-\mu_l\) for \((i,j)\in\Omega_l\),
and \(\Delta_y(i,j)=\mu_r-\mu_c\) for \((i,j)\in\Omega_r\).
The block mean difference terms such as $(\mu_c - \mu_t)\cdot\epsilon$ are precomputed before processing each block, and are stored in memory for quick access.

In HSZp-2d, derivatives can be computed directly from $D_p$ as
$
\nabla d_{i,j} = \Big((\sum_{s=i}^{i+1}\sum_{t=1}^j p_{s,t})\cdot\epsilon,\, (\sum_{s=1}^{i}\sum_{t=j}^{j+1} p_{s,t})\cdot\epsilon\Big)
$
which leads to the recursive formulation
\begin{equation}
    \vspace{-.2em}
    \begin{aligned}
        \delta_x d_{i,j} = \delta_x d_{i,j-1} + (p_{i,j}+p_{i+1,j})\cdot\epsilon,\\
        \delta_y d_{i,j} = \delta_y d_{i-1,j} + (p_{i,j}+p_{i,j+1})\cdot\epsilon.
    \end{aligned}
    \label{eq:derivative_Dp}
    \vspace{-.2em}
\end{equation}
We compute derivatives row by row in a bottom-up order. For $\delta_xd_{i,j}$, when processing row $i$, we maintain a scalar accumulator $\eta$, initialized to zero at $j=1$, that tracks the prefix sum $\eta = \sum_{t=1}^j (p_{i,t} + p_{i+1,t})\cdot\epsilon$. At each column $j$, we update $\eta$ and compute $\delta_xd_{i,j} = \delta_xd_{i,j-1} + \eta$. Similarly, for $\delta_yd_{i,j}$, we maintain a column-wise accumulator $\eta_j = \sum_{s=1}^i (p_{s,j}+p_{s,j+1})\cdot\epsilon$ which is updated once per row.

\circled{3}\textit{ Operation with $D_q$}: From Eq.~\ref{eq:derivatives}, computing $\nabla d_{i,j}$ from $D_q$ follows the same procedure as from $D_f$, the only difference is the point at which dequantization is applied. Since $d'_{i,j} = q_{i,j}\cdot2\epsilon$, the derivative vector is given by
\vspace{-.5em}
\begin{equation}
    \delta_x d_{i,j} = [q_{i+1,j} - q_{i-1,j}]\cdot\epsilon,\;
    \delta_y d_{i,j} = [q_{i,j+1} - q_{i,j-1}]\cdot\epsilon
\end{equation}

\subsubsection{Laplacians} Homomorphic Laplacian computations can be completed in stages \circled{2}\circled{3} for multidimensional compressors and in stage \circled{3} for 1D compressors.

\circled{2}\textit{ Operation with $D_p$}: 
In HSZx-2d, for an interior point $(i,j)$ within a block, we first compute $\eta = (p_{i-1,j} + p_{i+1,j} + p_{i,j-1} + p_{i,j+1} - 4p_{i,j})\cdot2\epsilon$. For boundary points, $\eta$ is corrected by adding $(\mu_l-\mu_c)\cdot2\epsilon$ if $(i,j)\in\Omega_l$, $(\mu_r-\mu_c)\cdot2\epsilon$ if $(i,j)\in\Omega_r$, $(\mu_t-\mu_c)\cdot2\epsilon$ if $(i,j)\in\Omega_t$ and $(\mu_b-\mu_c)\cdot2\epsilon$ if $(i,j)\in\Omega_b$. The resulting value gives $\Delta d_{i,j} = \eta$.

In HSZp-2d, Laplacians can be computed directly from $D_p$ by rewriting Eq.~\ref{eq:laplacian_Dq} as
\vspace{-.2em}
    \begin{align} \notag
        \Delta d_{i,j} &= [-(q_{i,j}-q_{i-1,j})+(q_{i+1,j}-q_{i,j})\\ \notag
        &\qquad\qquad -(q_{i,j}-q_{i,j-1}) + (q_{i,j+1}-q_{i,j})]\cdot2\epsilon\\
        &= (-\xi_{11} + \xi_{12} -\xi_{2,j} + \xi_{2,j+1})\cdot2\epsilon.
    \end{align}
    \label{eq:laplacian_Dp}
Here $\xi_{11}$ and $\xi_{12}$ denote prefix sums along rows $i$ and $i+1$ up to column $j$, while $\xi_{2j}$ and $\xi_{2,j+1}$ store column-wise prefix sums along row $i$, each up to column $j$ and column $j+1$. During row-wise processing, $\xi_{11}$ and $\xi_{12}$ are reset at the start of each row and updated incrementally across columns, whereas $\xi_{2,*}$ is updated once per row.

\circled{3}\textit{ Operation with $D_q$}: 
Reviewing Eq.~\ref{eq:laplacian}, the formula for computing $\Delta d_{i,j}$ using $D_q$ is given below:
\vspace{-.5em}
\begin{equation}
    \Delta d_{i,j} = [q_{i-1,j} + q_{i+1,j} + q_{i,j-1} + q_{i,j+1} -  4q_{i,j}]\cdot2\epsilon.
    \label{eq:laplacian_Dq}
\end{equation}

\subsection{Multivariate Derivation}\label{sec:multivariate}
Since divergence and curl reduce to component-wise $\delta_x$ and $\delta_y$, we compute both operators using the same homomorphic methodology developed for scalar derivatives.

\subsubsection{Divergence}
Let $\{p^u_{i,j}\}$ and $\{p^v_{i,j}\}$ denote the decorrelated data of the two components of the vector field $\mathbf{F} = (\mathbf{u}, \mathbf{v})$. 
Homomorphic divergence can be computed in stages \circled{2}\circled{3} for multidimensional compressors, and in stage \circled{3} for 1D compressors.

\circled{2}\textit{ Operation with $D_p$}: 
In HSZx-2d, the divergence at interior positions of a block is computed as
\vspace{-.5em}
\begin{equation}
    (\nabla\!\cdot\!\mathbf{F})_{i,j} = [(p^u_{i+1,j} - p^u_{i-1,j}) + (p^v_{i,j+1} - p^v_{i,j-1})]\cdot\epsilon.
\end{equation}
For border positions of the block, the block mean difference terms of each vector field component are added to the formula above.
In HSZp-2d, we can directly apply the recursive formula Eq.~\ref{eq:derivative_Dp} in each computation of $\delta_x d_{i,j}$ and $\delta_y V_{i,j}$, then add up the results to produce the divergence.

\circled{3}\textit{ Operation with $D_q$}: 
Let $\{q^u_{i,j}\}$ and $\{q^v_{i,j}\}$ denote the quantized data of $U$ and $V$, respectively. The formula for computing divergence using $D_q$ is
\vspace{-.5em}
\begin{equation}
    (\nabla\!\cdot\!\mathbf{F})_{i,j} = [(q^u_{i+1,j} - q^u_{i-1,j}) + (q^v_{i,j+1} - q^v_{i,j-1})]\cdot\epsilon.
    \vspace{-.2em}
\end{equation}

\subsubsection{Curl} Homomorphic curl computation can also be completed in stages \circled{2}\circled{3} for multidimensional compressors and in stage \circled{3} for 1D compressors.

\circled{2}\textit{ Operation with $D_p$}: 
In HSZx-2d, we use the same method discussed in derivative operation to compute $\delta_y d_{i,j}$ and $\delta_x V_{i,j}$;
in HSZp-2d, we follow Eq.~\ref{eq:derivative_Dp} to obtain these two fragments. Then $(\nabla\!\times\!\mathbf{F})_{i,j}$ equals $\delta_y d_{i,j} - \delta_x V_{i,j}$.

\circled{3}\textit{ Operation with $D_q$}: 
Similar to divergence, the computation of curl using $D_q$ follows
\vspace{-.5em}
\begin{equation}
    (\nabla\!\times\!\mathbf{F})_{i,j} = [(q^u_{i,j+1} - q^u_{i,j-1})- (q^v_{i+1,j} - q^v_{i-1,j})]\cdot\epsilon.
\end{equation}

\subsection{Error Analysis}
In this subsection, we analyze the bias between operation results computed from partially decompressed data $D_m, D_p$, $D_q$ and those from $D_f$, proving that the bias terms are all bounded by $\epsilon$.
\subsubsection{Bias of Mean computed in $M$ in HSZx and HSZx-nd}
Computing $\mu$ from metadata $M$ incurs larger error, but it is still bounded by $\epsilon$. For each block $b$, $M_b = \texttt{round}\bigl(\frac{\sum_{q\in D_q[B_b]}q}{S_b}\bigr) = \frac{\sum_{q\in D_q[B_b]}q}{S_b} - r_b,\; |r_b|\le \frac{1}{2}$. Using Eq.~\ref{eq:mean_m}, we have 
$
\sum_{b=1}^{n_b}M_bS_b = \sum_{q\in D_q}q - \sum_{b=1}^{n_b}r_bS_b,\; \big|\sum_{b=1}^{n_b}r_bS_b\big|\le \frac{1}{2}N.
$
Thus
$
\mu_M = \frac{1}{N}\bigl(\sum_{q\in D_q}q - \sum_{b=1}^{n_b}r_bS_b\bigr)\cdot2\epsilon,\, \mu_f = \frac{1}{N}\sum_{q\in D_q}(q\cdot 2\epsilon),
$
and the bias satisfies
$
|\mu_M - \mu_f| \le \frac{1}{N}\Big|\sum_{b=1}^{n_b}r_bS_b\Big|\cdot2\epsilon \le \epsilon.
$
\subsubsection{Bias of Mean computed in $D_p$ and $D_q$ approaches in all compressors}\label{sec:bias_mean_simple}
For both HSZp and HSZx, mean computation using $D_p$ or $D_q$ accumulates integer sums before scaling, whereas using $D_f$ evaluates the same expression in a different order, so the only source of bias is floating point rounding, which is $O(e\cdot\epsilon)$ and negligible compared to $\epsilon$.
\subsubsection{Bias of standard deviation computed in HSZx-$D_p$ and HSZx-nd-$D_p$}\label{sec:bias_bar_single}
For any compressor, the standard deviation computed from $D_f$ is
$
\sigma_f
= \{\sum_i(q_i - \frac{\sum_{i'} q_{i'}}{N})^2 / (N-1)\}^{\frac{1}{2}} \cdot 2\epsilon.
$
In the $D_p$ approach, the mean is
$
\tilde{\mu}
= \frac{1}{N}\left(\sum_{q\in D_q} q - \sum_{b=1}^{n_b} r_b S_b\right),
$
where $|\sum_b r_b S_b|/N = |\tilde{r}| \le \frac{1}{2}$. The resulting standard deviation is 
$
\sigma_p
= \{\sum_i\left(q_i - \tilde{\mu}\right)^2/(N-1)\}^{\frac{1}{2}}\cdot2\epsilon
= \{\sum_i\left(q_i - \frac{1}{N}\sum_{i'} q_{i'} + \tilde r\right)^2/(N-1)\}^{\frac{1}{2}}\cdot2\epsilon.
$
Defining $x_i = q_i - \frac{1}{N}\sum_{i'} q_{i'}$ and using reverse triangle inequality of the Euclidean norm, the bias reduces to
$
\beta =
|(\sum_i x_i^2)^{\frac{1}{2}} - (\sum_i (x_i+\tilde r)^2)^{\frac{1}{2}}|/(N-1)^{\frac{1}{2}}
\le \sqrt{N}|\tilde r|/\sqrt{N-1} \sim |\tilde r| \le\frac{1}{2}.
$
Hence $|\sigma_f - \sigma_p|\le\beta\cdot2\epsilon\le\epsilon$.
\subsubsection{Bias of standard deviation computed in all other approaches}
Although $\sum_{q\in D_q} q$ and $\sum_{q\in D_q} q^2$ are computed differently across the three approaches, all results reduce to Eq.~\ref{eq:var_q}, which is mathematically identical to $\sigma_f$, so the only source of bias is floating point rounding.
\subsubsection{Pointwise Bias in numerical differentiation and multivariate derivation}
For both operations using $D_q$ or $D_q$, finite difference operators are computed exactly in integer form and then scaled, making the results mathematically identical to those from $D_f$, thus the only source of bias is round-off errors in floating point arithmetic.

%% file: tex/evaluation.tex
\section{evaluation}\label{sec:evaluation}
In this section, we evaluate our methods in terms of compression ratios, decompression throughput, and analytical operation throughput using five real-world datasets from scientific simulations. 

\subsection{Experimental Setup}
Our benchmark datasets span multiple computational domains, with specifications summarized in \cref{tab:data}. 
All data are stored in single precision floating point format. We use all fields of the five datasets to evaluate statistical computations and numerical differentiation. For multivariate derivation, we use the velocity vector field from each dataset.
All experiments are conducted on a single core of a medium scale computing cluster~\cite{mcc}, where each node is equipped with two AMD EPYC Rome 7702P processors and 512 GB of memory.

\begin{table}[ht]
{
\footnotesize
\centering
\vspace{-1em}
\caption{Benchmark datasets.}
\vspace{-1em}
\label{tab:data}
\footnotesize
\resizebox{\columnwidth}{!}{%
\begin{tabular}{|l|c|c|c|c|c|}
\hline
\thead{Dataset} & \thead{\#Field} & \thead{Dimensions} & \thead{Size} & \thead{Domain}\\ 
\hline
Ocean & 2 & $2400\times 3600$ &  65.91 MB & Oceanology \\
\hline
Miranda & 7 & $256\times 384\times 384$ & 0.98 GB & Hydrodynamics\\
\hline
Hurricane & 13 & $100\times 500\times 500$ &  1.21GB & Climate \\
\hline
NYX & 6 & $512\times 512\times 512$ & 3.00 GB & Cosmology\\
\hline
JHTDB & 3 & $2580\times 2580\times 2580$ &  191.92 GB & Turbulence\\
\hline
\end{tabular}
}
\vspace{-1em}
}
\end{table}

We evaluate the effectiveness of the proposed methods using aggregated \textit{compression ratios} and \textit{operation throughput}. 
The former is measured by the total size of input data over that of compressed data, indicating how much data is reduced under a user-specified error tolerance. 
The latter is computed by the total size of input data over the total elapsed operation execution time, demonstrating how fast the operation is performed. 
Higher numbers in both metrics represent better efficiency. 

\subsection{Compression Ratios and Decompression Throughput}
\begin{figure}[ht]
	\vspace{-1.5em}
	\includegraphics[width=0.95\columnwidth]{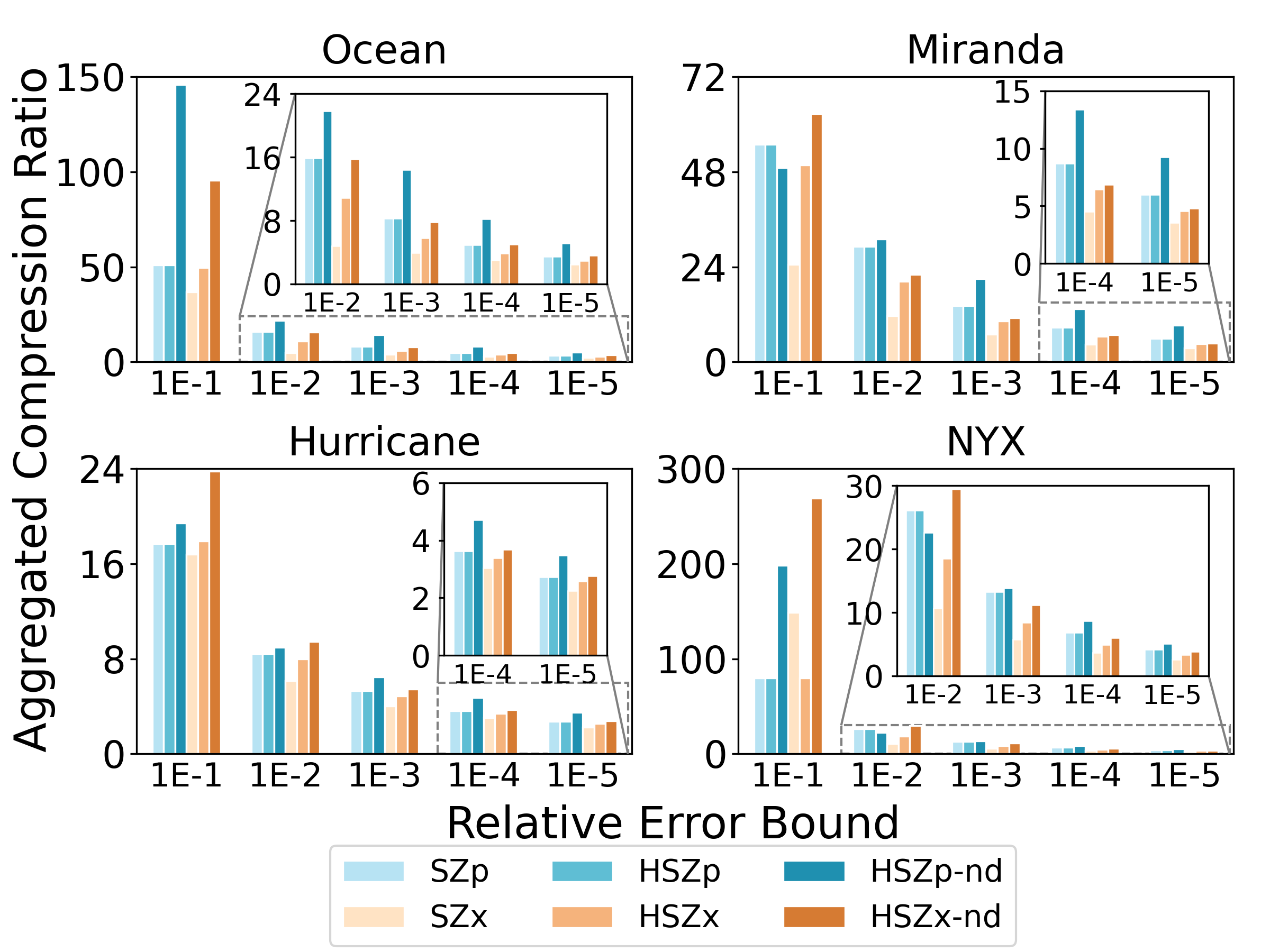}
	\vspace{-1em}
	\centering
	\caption{Compression ratios of homomorphic compressors and baselines.}\label{fig:cr}
\end{figure}

We first compare the compression ratios of our compressors with their baseline counterparts, i.e., HSZp \& HSZp-nd versus SZp, and HSZx \& SZx-nd versus SZx. Figure~\ref{fig:cr} presents the aggregated compression ratios under five value-range-based relative error bounds. 
Note that HSZp always has similar compression ratios to SZp because they share a very similar compression workflow.
According to the figure, HSZx consistently achieves higher compression ratios than SZx, particularly under tighter error bounds. This improvement stems from HSZx’s use of blockwise mean values, which smooth the decorrelated data distribution and reduce the encoded size. The figure also shows that HSZp-nd and HSZx-nd significantly improve the compression ratios of SZp and SZx across most error bounds, reinforcing our decision to extend HSZp and HSZx to support multidimensional data.

\begin{figure}[ht]
	\includegraphics[width=0.95\columnwidth]{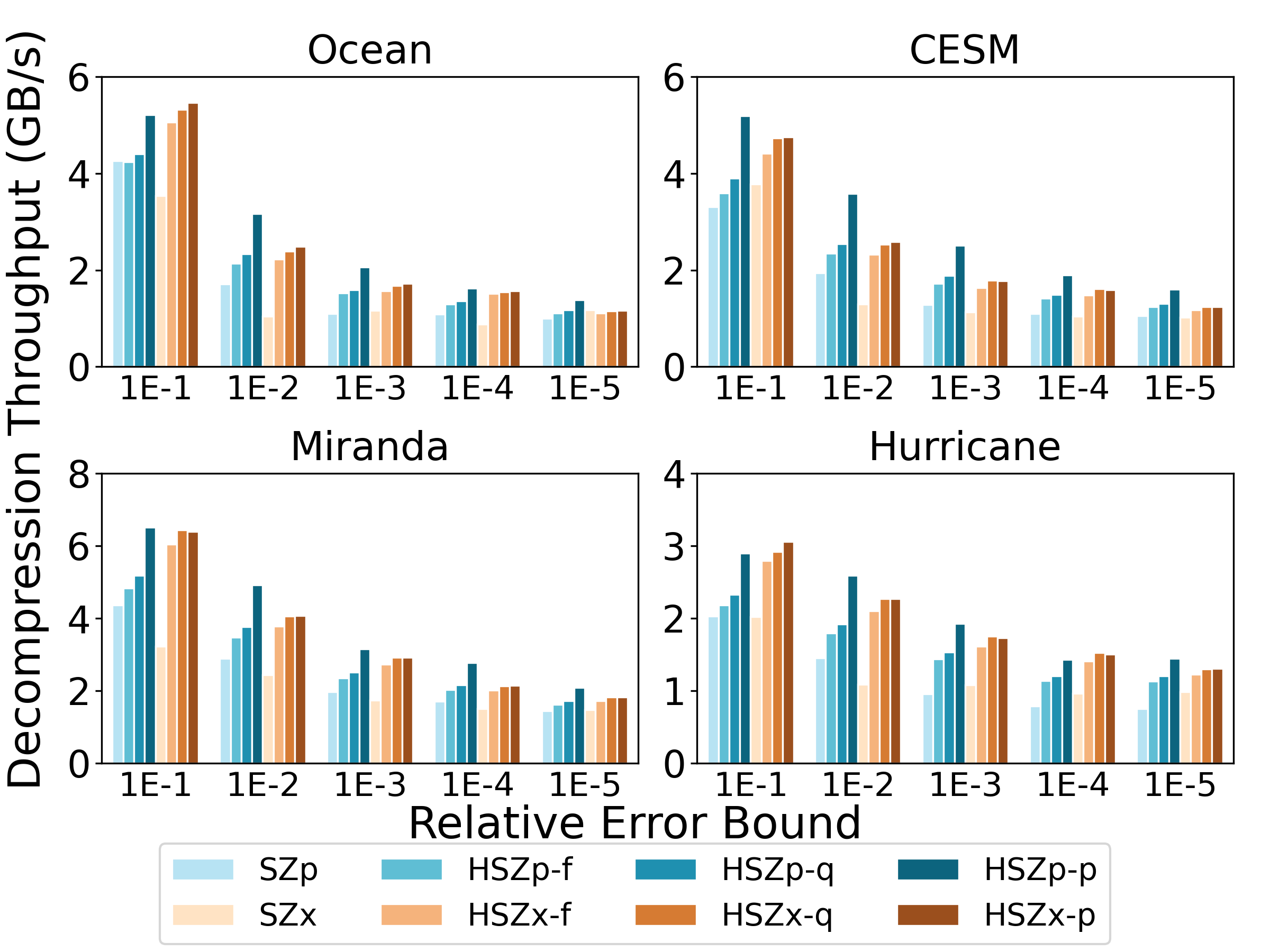}
	\vspace{-1em}
	\centering
	\caption{Decompression throughput of 1D compressors.}\label{fig:decompress}
	\includegraphics[width=0.95\columnwidth]{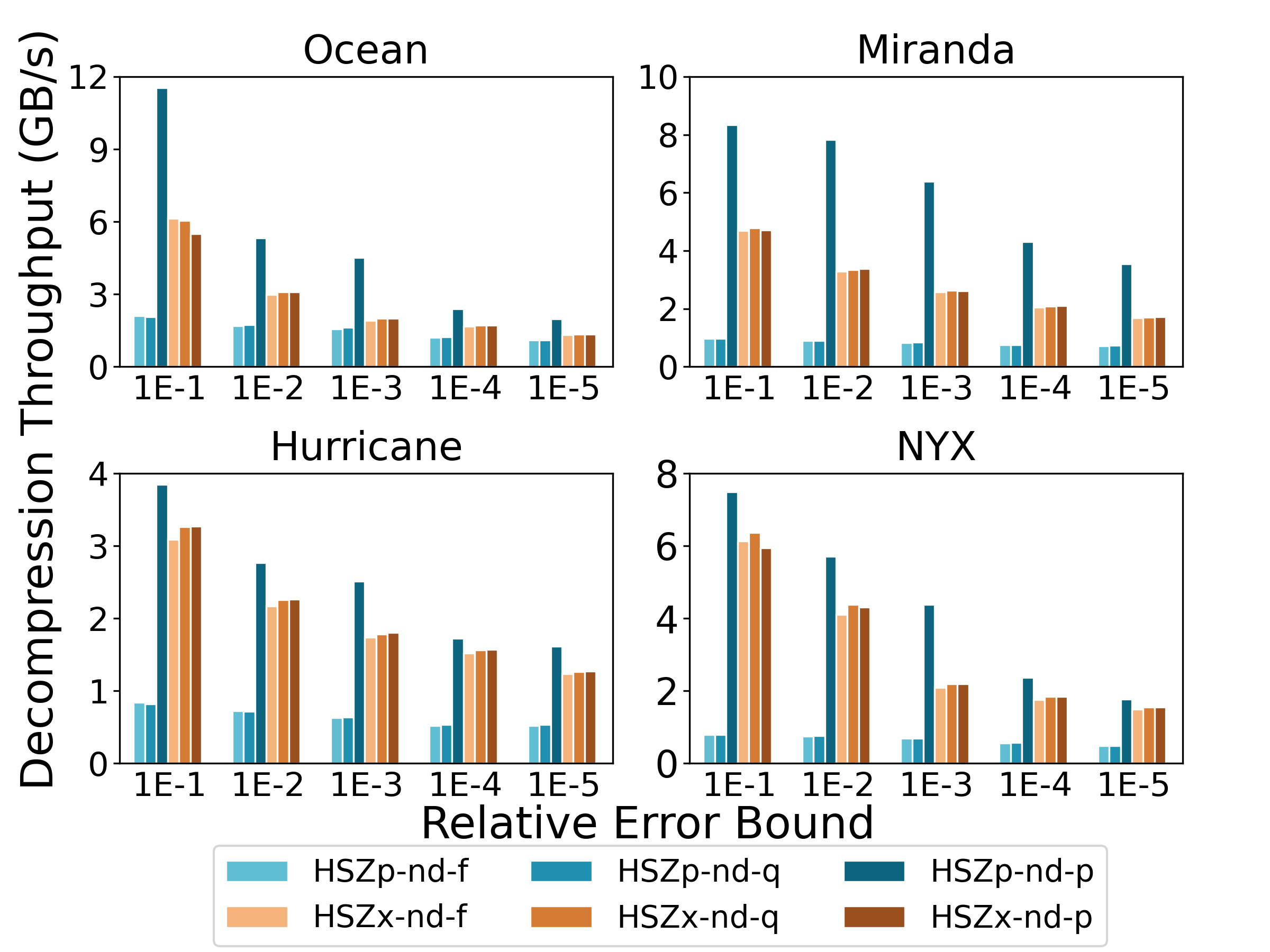}
	\vspace{-1em}
	\centering
	\caption{Decompression throughput of ND compressors.}\label{fig:decompress-nd}
	\vspace{-.5em}
\end{figure}

In addition to the compression ratio improvement, HSZp and HSZx are faster in decompression than SZp and SZx. In \cref{fig:decompress}, SZp, SZx, HSZp-f, and HSZx-f denote full decompression, while partial decompression to $D_q$ and $D_p$ are represented by –q and –p, respectively.
The figure depicts a stepwise improvement in partial decompression performance between stage \circled{2} and stage \circled{3} for both HSZp and HSZx. 
Stepwise improvement in HSZx is less noticeable than that in HSZp, because HSZx uses a comparatively simple prediction mechanism, which leads to similarly high partial decompression performance in both HSZx-q and HSZx-p.
\cref{fig:decompress-nd} shows that HSZp-nd-p achieves significantly higher decompression throughput than HSZp-nd-q and HSZp-nd-f, whereas HSZx-nd exhibits little speedup or degradation across decompression stages under different error bounds. This behavior can be explained by reasons similar to those discussed for HSZx.
\vspace{-1em}

\subsection{Analytical Operation Throughput}
We then present the throughput of three categories of analytical operations on scientific data of all evaluated methods. 
Since the trends between 1D compressors and ND compressors are similar across different analytic operations, we omit the results of 1D compressors for numerical differentiation and multivariate derivation due to limited space. 

\begin{figure}[t]
	\vspace{-1em}
	\includegraphics[width=0.92\columnwidth]{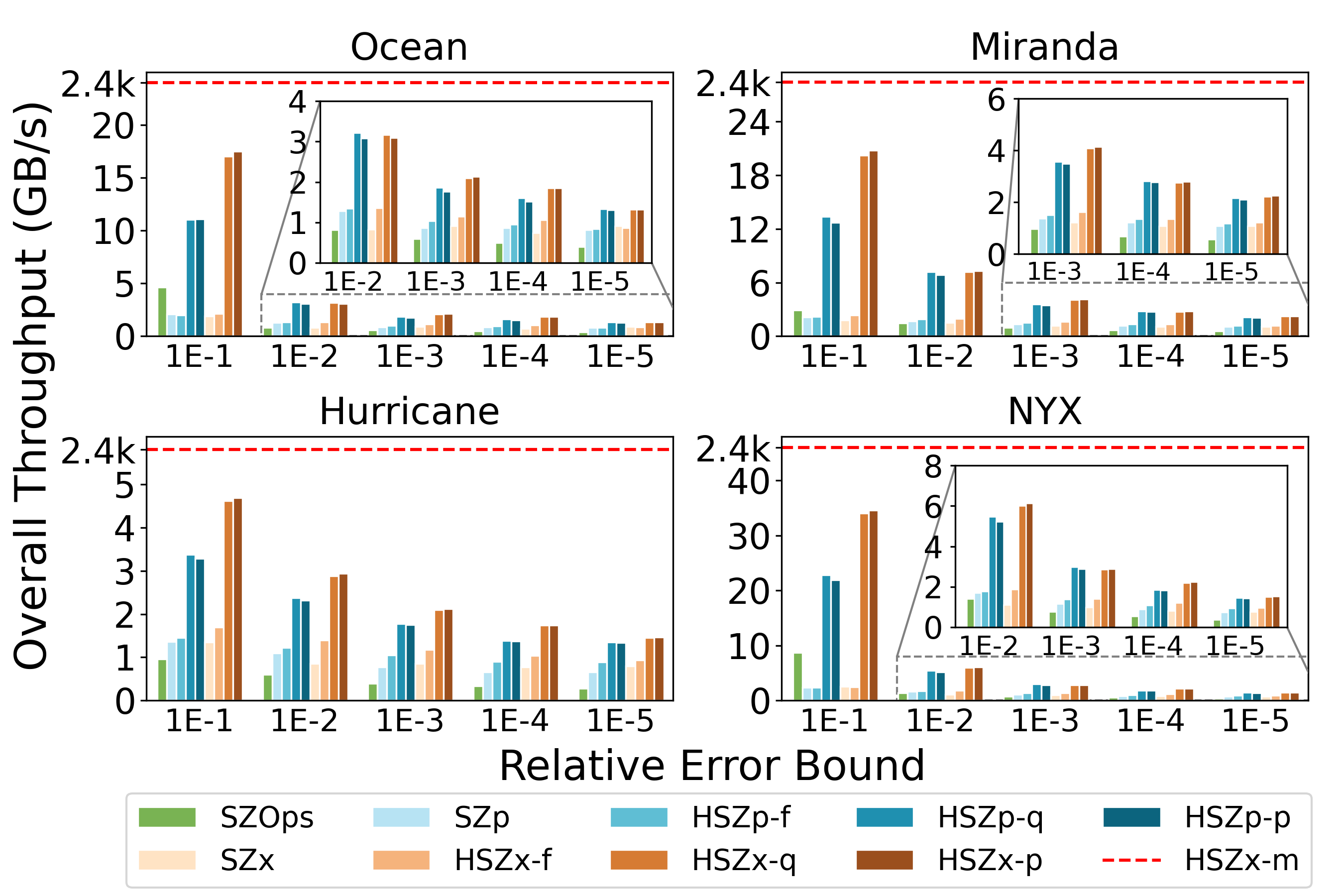}
	\centering
	\vspace{-1em}
	\caption{Throughput of computing mean (1D).}\label{fig:mean}
	\includegraphics[width=0.92\columnwidth]{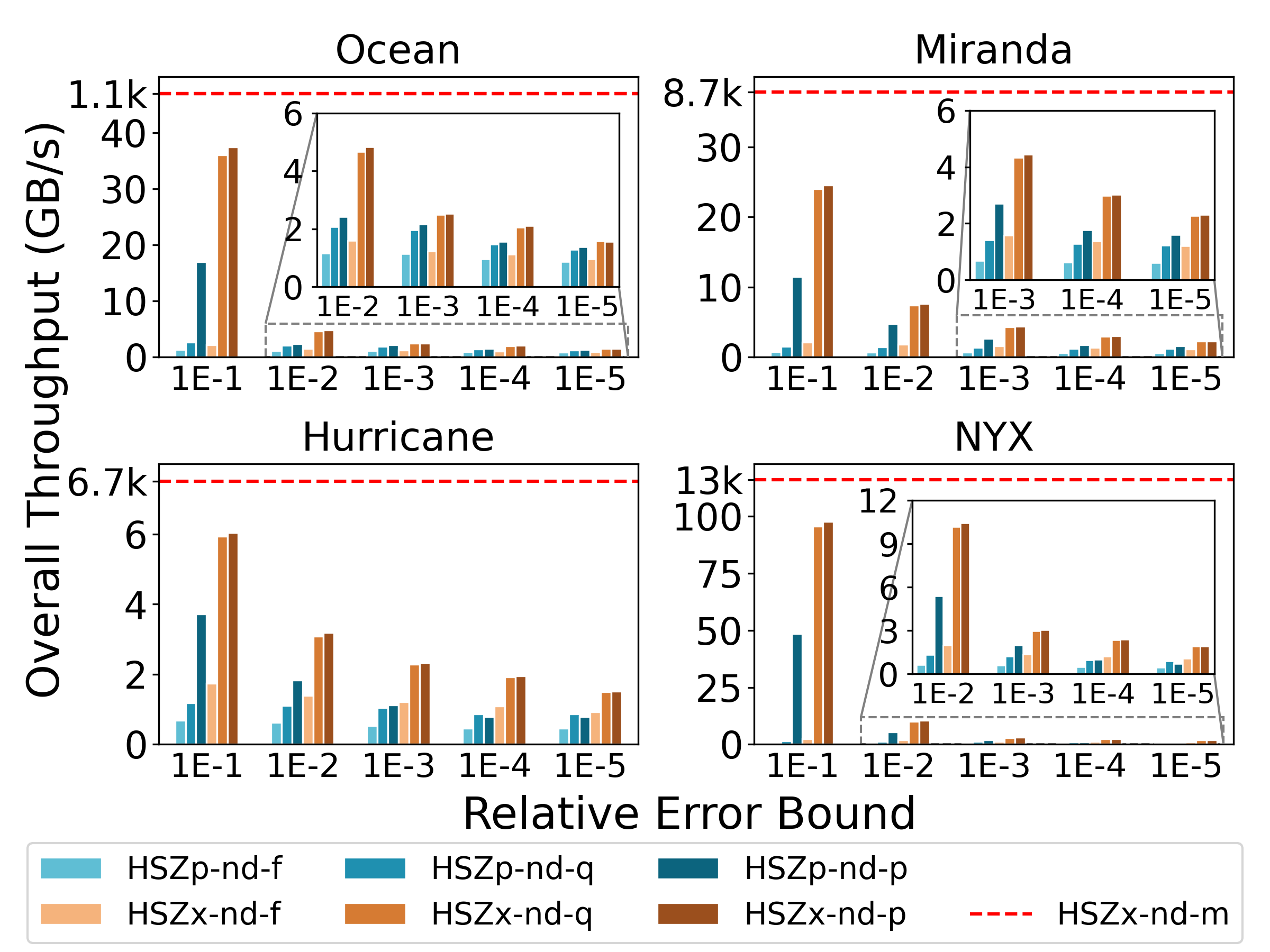}
	\centering
	\vspace{-1em}
	\caption{Throughput of computing mean (ND).}\label{fig:mean-nd}
	\vspace{-.5em}
\end{figure}

\subsubsection{Statistical Computation}
\begin{figure}[t]
	\vspace{-1em}
	\includegraphics[width=0.9\columnwidth]{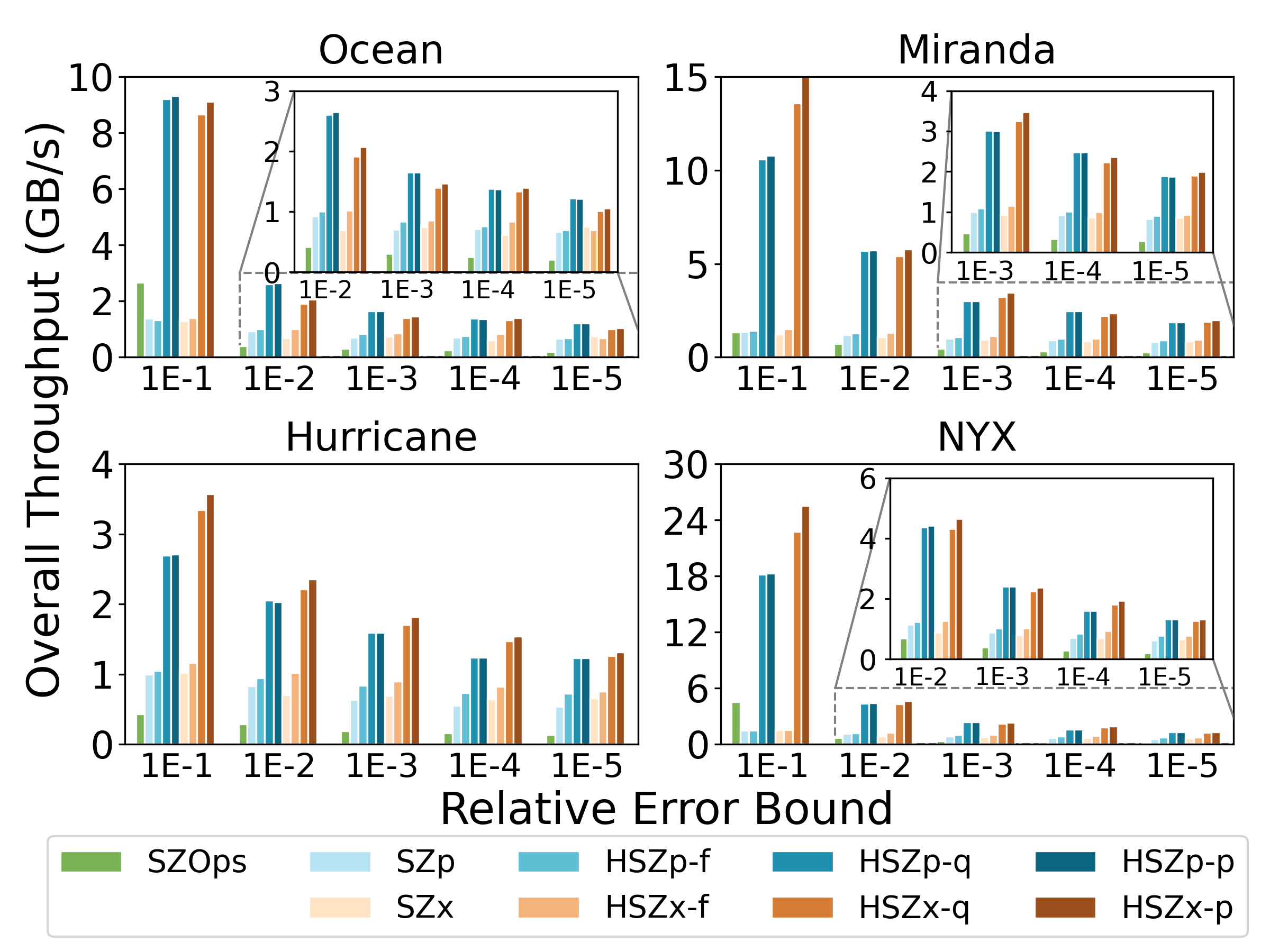}
	\centering
	\vspace{-1em}
	\caption{Throughput of computing standard deviation (1D).}\label{fig:std}
	\includegraphics[width=0.9\columnwidth]{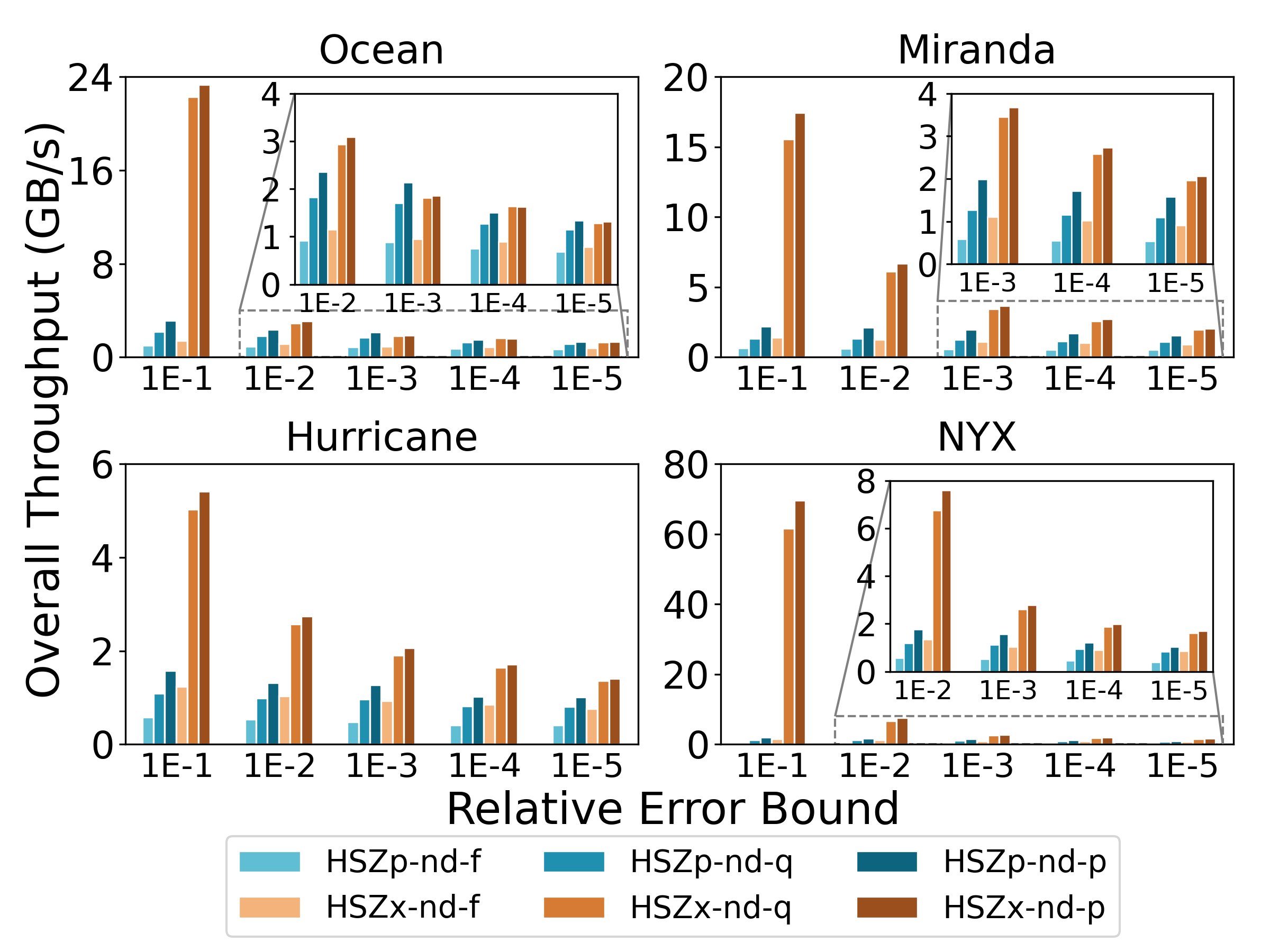}
	\centering
	\vspace{-1em}
	\caption{Throughput of computing standard deviation (ND).}\label{fig:std-nd}
\end{figure}

We compare the operation throughput of HSZp and HSZx against SZp, SZx, and SZOps. The results for mean and standard deviation are presented in Figure~\ref{fig:mean} and Figure~\ref{fig:std}, respectively. The red dashed line in Figure~\ref{fig:mean} (and Figure~\ref{fig:mean-nd}) indicates the minimum throughput of HSZx-m across all error bounds. 
SZOps has higher throughput than all the generic workflow at a loose error bound of 1E-1 because it performs homomorphic operations on $D_q$, but all of our homomorphic approaches greatly outpace SZOps under all error bounds.
HSZp-p, HSZp-q, HSZx-p, HSZx-q, and HSZx-m exhibit high overall throughput for both mean and standard deviation operations, with HSZx-m marking the highest throughput of over 2400GB/s in computing the mean under all error bounds. 
For standard deviation computation with a relative error bound of 1E-1, HSZx-p achieves 3.49$\times$, 7.99$\times$, 6.20$\times$, and 4.04$\times$ higher throughput than the best existing approach across the four datasets.
Figure~\ref{fig:mean-nd} and Figure~\ref{fig:std-nd} report results for ND compressors, where HSZx-nd has even larger advantage over HSZp-nd than in 1D compressors. At a relative error bound of 1E-1, HSZx-nd-m boosts mean value operation throughput by 263$\times$, 3079$\times$, 7315$\times$, and 1676$\times$ over SZOps on the four datasets.
In summary, HSZx (and HSZx-nd) offer superior performance in statistical operations.
Since the 1D and ND compressors exhibit similar patterns across different decompression stages, we report only the ND compressor results for the following operations.

\subsubsection{Numerical Differentiation}
\begin{figure}[ht]
	\includegraphics[width=0.92\columnwidth]{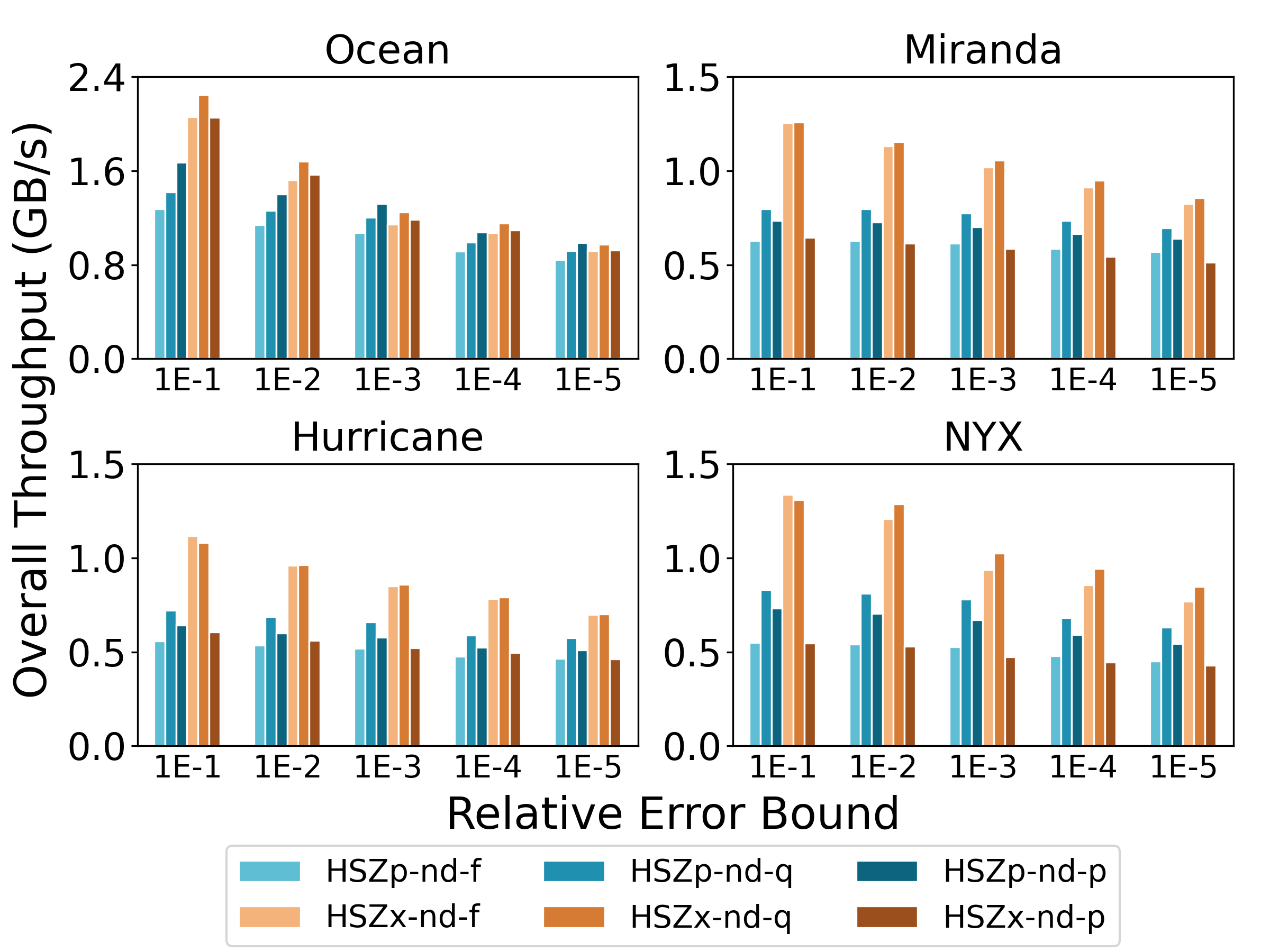}
	\centering
	\vspace{-1em}
	\caption{Throughput of computing derivatives (ND).}\label{fig:derivatives-nd}
    \vspace{-1em}
\end{figure}

\begin{figure}[t]
	\includegraphics[width=0.92\columnwidth]{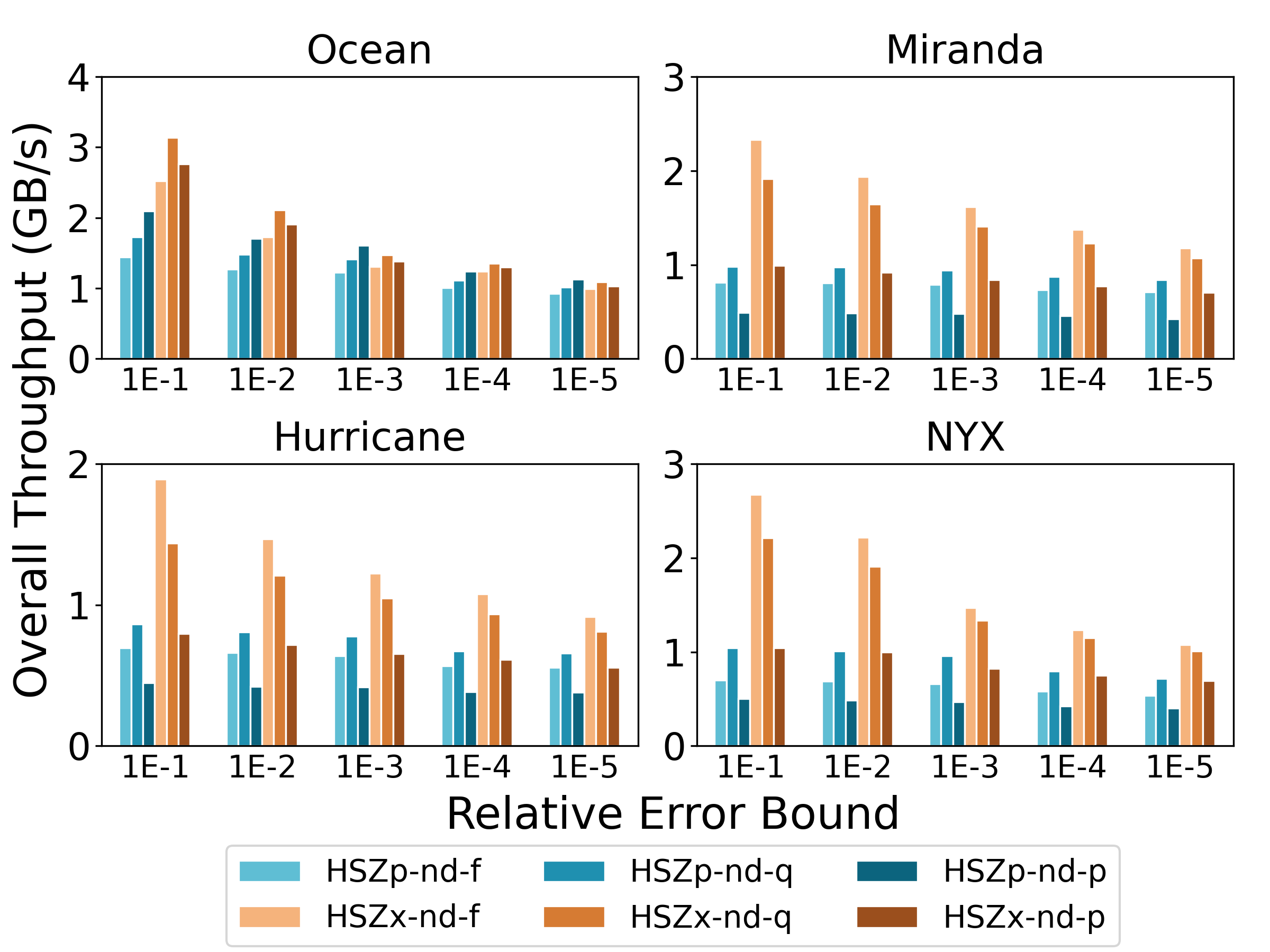}
	\centering
	\vspace{-1em}
	\caption{Throughput of computing Laplacians (ND).}\label{fig:laplacians-nd}
\end{figure}

\cref{fig:derivatives-nd} and \cref{fig:laplacians-nd} present results for derivative and Laplacian operations using ND compressors. In the 2D setting, HSZp-nd shows clear stepwise speedups from HSZp-nd-f to HSZp-nd-q and HSZp-nd-p. For derivative computation on the Ocean dataset at a relative error bound of 1E-1, HSZp-nd-p reaches 1.11$\times$ the throughput of HSZp-nd-q, and HSZp-nd-q reaches 1.18$\times$ that of HSZp-nd-f, yielding an overall 1.31$\times$ speedup over the generic HSZp-nd-f workflow.
In the 3D setting, the $D_p$ approaches generally underperform the $D_q$ approaches. HSZx-nd-p is marked with the most degradation, as HSZx-3d incurs heavier block border processing overhead than HSZx-2d.
Nevertheless, HSZp-nd-p consistently achieves at least 1.10$\times$ the throughput of HSZp-nd-f across all 3D datasets and error bounds in derivative operations.
This improvement is enabled by the recursive design of HSZp-nd-p, which reduces derivative computation to a simple 1D prefix-sum.
In 3D case, derivatives along each dimension correspond to 2D Lorenzo prediction and can therefore be efficiently computed via fast 1D prefix-sum operations, as introduced in \cref{alg:var_p_2d}.


\begin{figure}[ht]
	\vspace{-1em}
	\includegraphics[width=0.92\columnwidth]{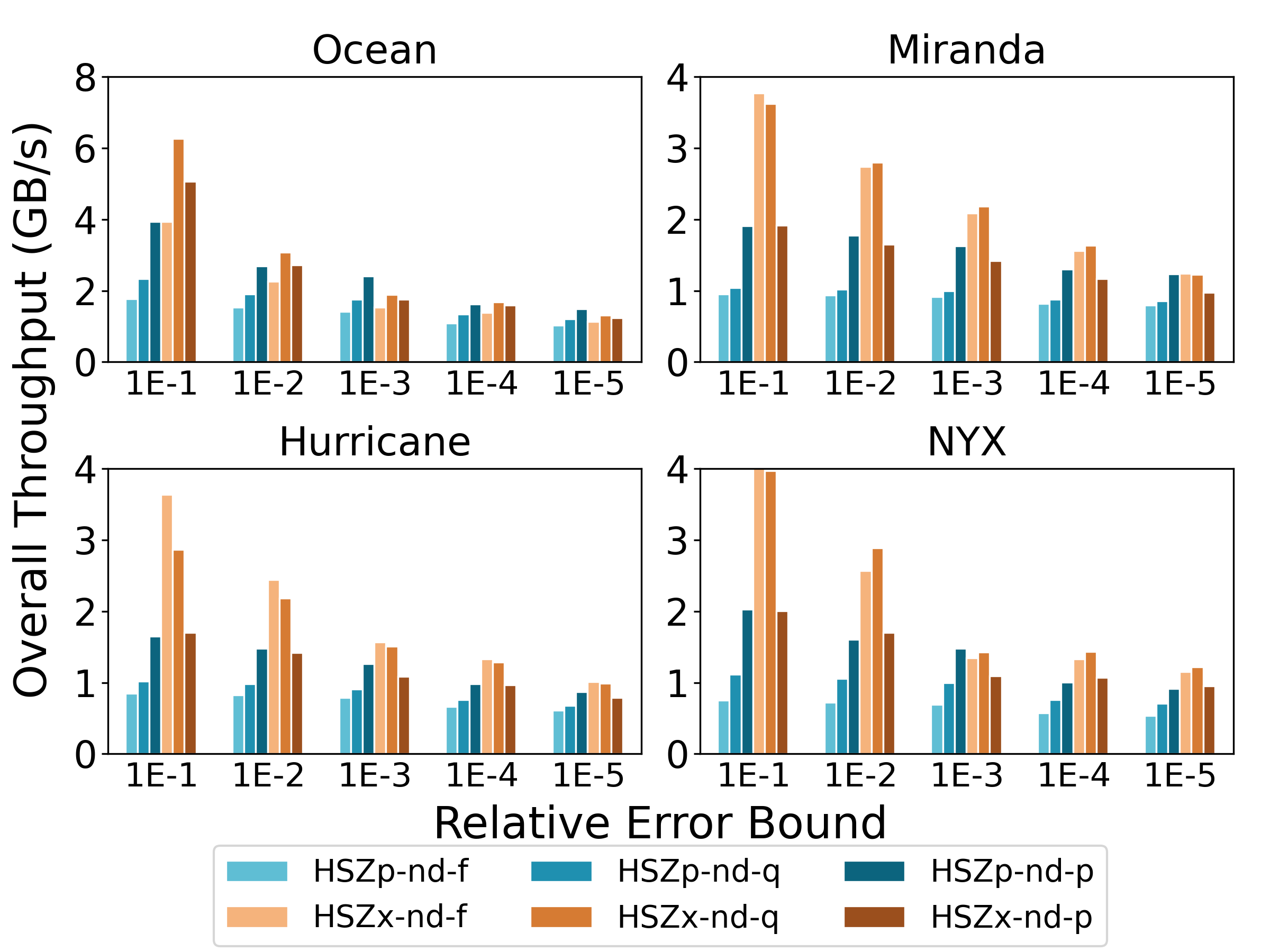}
	\centering
	\vspace{-1em}
	\caption{Throughput of computing divergence (ND).}\label{fig:divergence-nd}
    \vspace{-1em}
\end{figure}

\subsubsection{Multivariate Derivation}
When computing divergence, although only a single output is produced per data point, one partial derivative must be evaluated along each coordinate direction, resulting in two derivatives in 2D and three in 3D. Computing curl in 3D is more expensive, requiring six partial derivatives per data point from different component fields.
the performance bottleneck of HSZx-nd-p caused by block-boundary processing persists in 3D multivariate derivations. In contrast, HSZp-nd exhibits clear stepwise throughput improvements for divergence computation, as shown in \cref{fig:divergence-nd}. With a relative error bound of 1E-1, HSZp-nd-p achieves up to 2.00$\times$ the throughput of the baseline HSZp-nd-f across all datasets.
For curl computation, HSZp-nd-p demonstrates notable gains over HSZp-nd-f on 2D data, while HSZp-nd-q consistently outperforms HSZp-nd-f on both 2D and 3D datasets. In absolute terms, HSZp-nd-p maintains 2.22$\times$, 2.01$\times$, 1.94$\times$, and 2.68$\times$ the throughput of HSZp-nd-f in computing the divergence under error bound 1E-1.

\begin{figure}[t]
	\includegraphics[width=0.92\columnwidth]{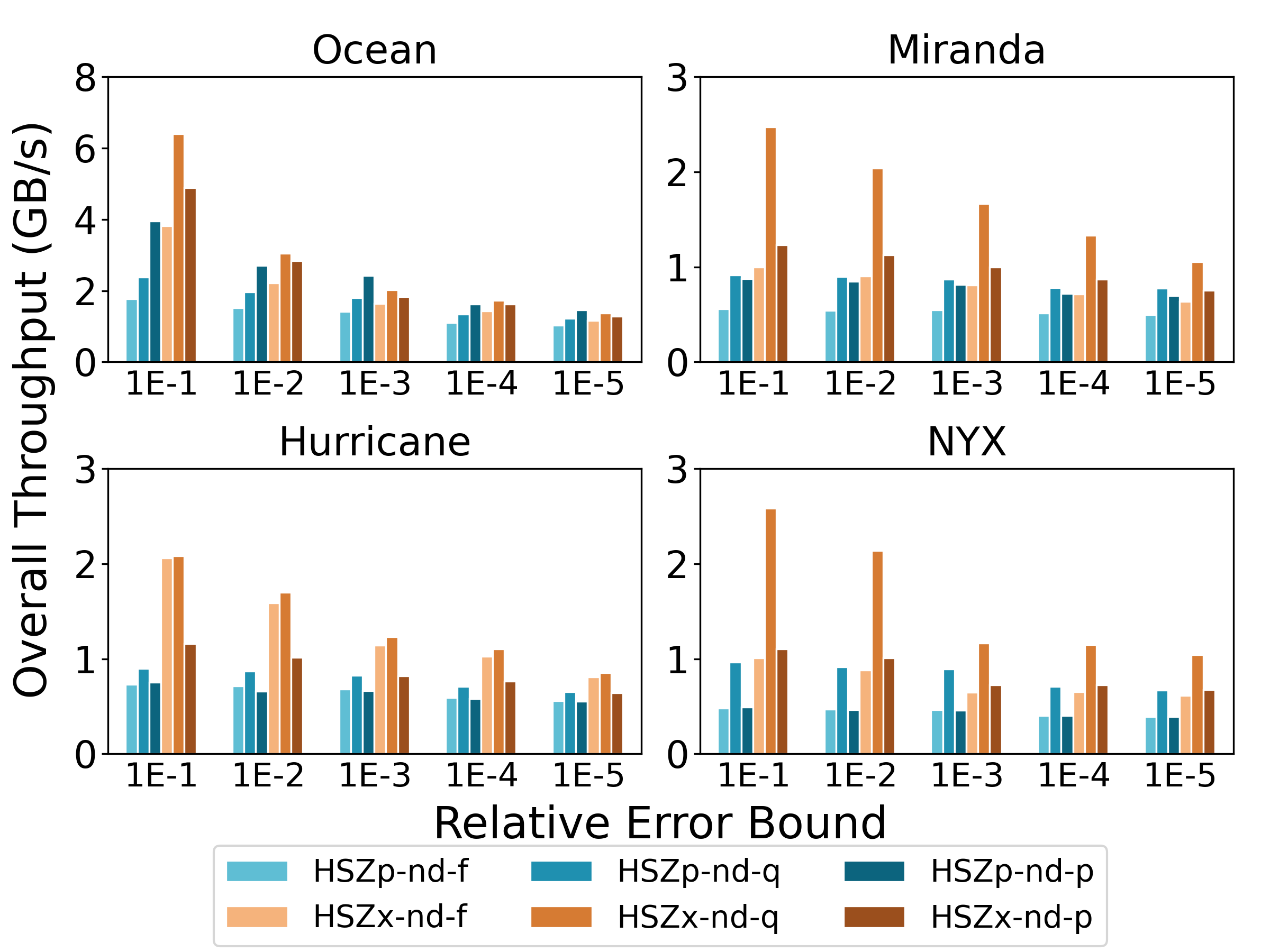}
	\centering
	\vspace{-1em}
	\caption{Throughput of computing curl (ND).}\label{fig:curl-nd}
\end{figure}
\begin{table}[ht]
{
\scriptsize
\centering
\vspace{-2em}
\caption{Time breakdown (in seconds)}
\vspace{-1em}
\label{tab:breakdown}
\footnotesize
\resizebox{\columnwidth}{!}{%
\begin{tabular}{|l|c|c|c|}
\hline
 & Decompression time & Computation time & Overall time\\ 
\hline
$D_f$ & 3.94 & 1.73 & 5.67\\
\hline
$D_q$ & 2.21 & 1.55 & 3.84\\
\hline
$D_p$ & 0.73 & 3.68 & 4.47\\
\hline
\end{tabular}
}
}
\vspace{-.5em}
\end{table}

\subsubsection{Complexity analysis for finite-difference-related operations}
When evaluating numerical differentiation and multivariate derivations, we observe that the $D_q$ approach can achieve higher throughput than $D_p$ in some cases. To explain this behavior, we use a 3D derivative computation on NYX dataset using HSZp-nd with a relative error bound of 1E-3, reporting a detailed breakdown of decompression and computation time in \cref{tab:breakdown}.
Both $D_p$ and $D_q$ have $O(n)$ decompression time and space complexity, but $D_q$ incurs an extra $O(1)$ cost per point and an additional $O(n)$ buffer to reconstruct quantization indices, making decompression 1.48 seconds slower than $D_p$ in this case. In contrast, derivative computation on $D_p$ requires auxiliary $O(n)$ buffers and extra $O(n)$ memory accesses to recover cross dimensional neighbors, increasing sensitivity to cache and bandwidth and causing a 2.13 second slowdown relative to $D_q$, which outweighs the decompression advantage and yields higher end to end throughput for $D_q$.

We observe the same trade-off for HSZx-nd. Because HSZx-nd relies more heavily on intermediate buffering at the $D_p$ stage, the computation-time penalty is even larger, making the crossover in favor of $D_q$ more pronounced in some cases.

\subsubsection{Analytical operation throughput on large datasets}
\begin{figure}[ht]
	\vspace{-1em}
	\includegraphics[width=\columnwidth]{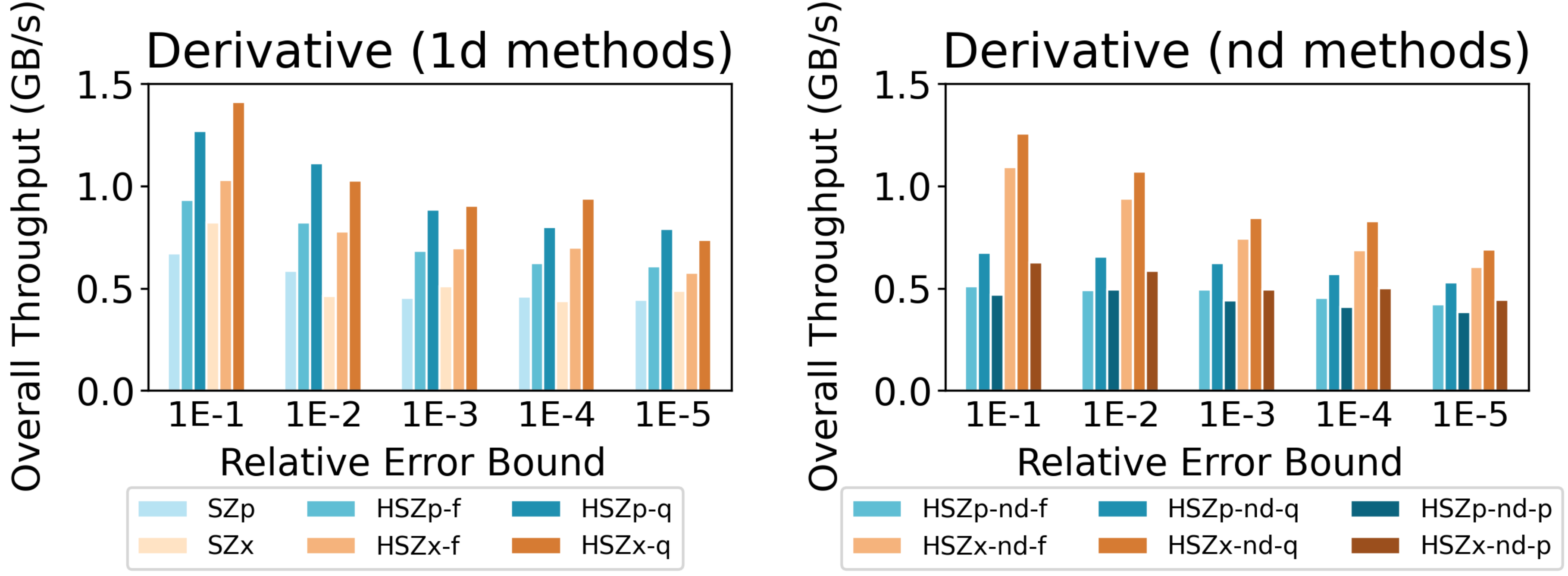}
	\centering
	\vspace{-2em}
	\caption{Throughput of computing derivatives using JHTDB dataset with 1D and ND compressors.}\label{fig:jhtdb}
	\vspace{-0.9em}
\end{figure}

To assess performance robustness on large scale data, we evaluate all analytical operations on the JHTDB dataset, which exceeds 190 GB. Due to space limitations, we report derivative throughput as a representative metric, since all six operations exhibit trends consistent with those observed on smaller datasets. As shown in \cref{fig:jhtdb}, under a relative error bound of 1E-1, HSZp-q achieves 1.89$\times$ the throughput of SZp and HSZx-q achieves 1.71$\times$ that of SZx. Likewise, HSZp-nd-q outperforms HSZp-nd-f by 1.31$\times$, and HSZx-nd-q exceeds HSZx-nd-f by 1.15$\times$ under the same error bound.

\subsubsection{Multi-operation throughput}
\begin{figure}[ht]
	\vspace{-1.2em}
	\includegraphics[width=0.95\columnwidth]{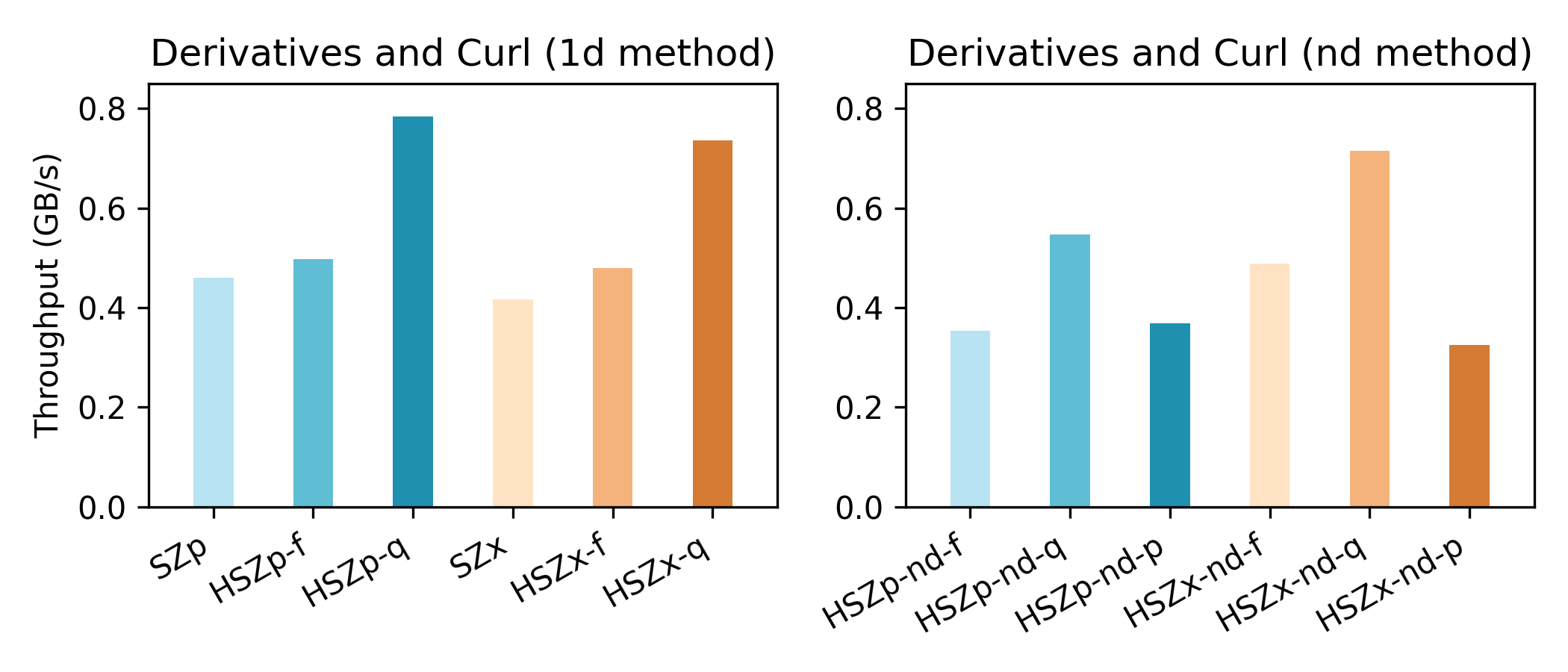}
	\centering
	\vspace{-1.5em}
	\caption{Throughput of computing derivatives and curl using NXY velocity vector field with 1D and ND compressors.}\label{fig:derivcurl}
	\vspace{-0.8em}
\end{figure}

We use the 3D velocity vector field from the NYX dataset to evaluate the “decompress once, run multiple operations” use case. This setting reflects typical NYX cosmological post processing.
The vector field contains three velocity components, and we compute both the derivatives of each velocity field and the curl of the resulting vector field. The experimental results shown in \cref{fig:derivcurl} demonstrate that our compressors consistently and significantly outperform the baselines. For 1D compressors, both HSZx and HSZp achieve higher throughput than SZx and SZp, respectively. Moreover, HSZx-q and HSZp-q each deliver over 1.50× speedup compared to their full-decompression counterparts (HSZx-f and HSZp-f). For 3D compressors, the $D_q$ approach similarly provides substantial benefits, achieving over 1.45× speedup compared to the $D_f$ baselines.

\subsubsection{Analytical operation errors}
We evaluate homomorphic operation errors by comparing results from the generic flow operating on fully decompressed data with those from homomorphic execution. \cref{tab:error} reports the maximum relative error across all six operations on the NYX dataset under a relative error bound of 1E-3, representing the worst case over all partial decompression stages. In all cases, the observed errors are well below the error bound, demonstrating that HSZ achieves both high accuracy and high performance.

\begin{table}[htb]
{
\footnotesize
\centering
\vspace{-1em}
\caption{Maximum relative error of six analytical operations in NYX dataset under relative error bound 1E-3.}
\label{tab:error}
\vspace{-1em}
\footnotesize
\resizebox{\columnwidth}{!}{%
\begin{tabular}{|l|c|c|c|c|c|c|}
\hline
& \thead{Mean} & \thead{\makecell{Standard\\ Deviation}} & \thead{Derivative} & \thead{Laplacian} & \thead{Divergence} & \thead{Curl} \\ 
\hline
HSZp & 9.03E-11 & 1.24E-10 & 5.22E-8 & 7.58E-7 & 3.74E-8 & 1.98E-8 \\
\hline
HSZx & 5.37E-5 & 4.44E-7 & 5.22E-8 & 7.58E-7 & 3.74E-8 & 1.98E-8 \\
\hline
HSZp-nd & 9.03E-11 & 1.24E-10 & 5.22E-8 & 7.58E-7 & 3.74E-8 & 1.98E-8 \\
\hline
HSZx-nd & 1.76E-5 & 4.44E-7 & 5.22E-8 & 7.58E-7 & 3.74E-8 & 2.10E-8 \\
\hline
\end{tabular}
}
}
\vspace{-.5em}
\end{table}

\vspace{-.5em}

\subsection{Compressor selection by use case}
Based on the experimental results, we recommend different compressors by use case. For statistical aggregations, HSZx-nd is preferred, as it consistently achieves the highest throughput among all methods. For 2D numerical differentiation and multivariate operations, HSZp-nd offers a better balance between compression ratio and end to end throughput. For 3D datasets, HSZx nd typically delivers higher throughput but at lower compression ratios. Accordingly, HSZx-nd is recommended when throughput is the primary concern, whereas HSZp-nd is preferable when higher compression ratios are desired.

\subsection{Discussion on generality}
In this work, we focus on SZp and SZx because they are representative high-throughput error-controlled compressors with prediction-based pipelines, which clearly expose the intermediate representations required by our multi-stage decompression design. Implementing HSZ on these compressors allows us to validate the key idea of performing analytical operations on intermediate representations while maintaining competitive compression performance.
Although our evaluation uses these two compressors as case studies, the proposed framework can generalize to other compressors because most scientific compression pipelines can be decomposed into multiple stages. However, the homomorphic operations on intermediate representations must be derived based on the underlying compressor and target tasks. For example, transform-based compressors may expose intermediate stages such as truncated transform coefficients, decompressed fixed-point integers, and decompressed floating-point numbers. If the Discrete Cosine Transform (DCT) is used, the mean can be recovered from the first truncated coefficient, whereas numerical differentiation typically requires decompressed fixed-point or floating-point values. Thus, while our framework is broadly compatible with different compression pipelines, the efficiency of homomorphic analytics depends on both the compressor and the target operation.

%% file: tex/conclusion.tex
\section{conclusion}\label{sec:conclusion}
In this paper, we present a novel framework to enable fast analytical operations on compressed scientific data. 
The key idea is to enable multi-stage decompression in the compression pipeline and design homomorphic analytic operations with intermediate data representations to avoid full decompression. 
We validate our framework by adapting two leading high-throughput scientific data compressors and their multidimensional variations. Experimental results demonstrate that the proposed homomorphic analytical operations deliver up to 7315$\times$, 1.89$\times$, and 2.68$\times$ speedup on three types of target analytical operations. 
In the future, we will investigate how to enable homomorphic analytical operations for more advanced scientific data compressors, support a wider range of analytical operations, explore adaptive stage-selection strategies, and extend the framework to multi-core CPU and GPU platforms to further improve throughput.